\documentclass[sigconf,10pt]{acmart}

\usepackage{subfigure}
\usepackage{mathrsfs}
\usepackage{amsfonts}
\usepackage{flushend}
\usepackage{amsmath}
\usepackage{latexsym}
\usepackage{paralist}
\usepackage{xspace}
\usepackage{multirow}
\usepackage{dsfont}
\usepackage{multicol}
\usepackage{etoolbox}
\usepackage{courier}
\usepackage{booktabs}
\usepackage{pifont}
\usepackage{setspace}
\usepackage{caption}
\usepackage{booktabs}
\usepackage{amscd}
\usepackage{relsize}
\usepackage{listings}
\usepackage{enumitem}
\usepackage{soul}
\usepackage{cancel}
\usepackage{textcase}
\usepackage{dblfloatfix}
\usepackage[flushleft]{threeparttable}
\usepackage{makecell}
\usepackage{etoolbox}
\usepackage{caption}  
\usepackage{booktabs}
\usepackage{float} 
\usepackage{mathtools}
\usepackage{tikz-cd}
\usepackage[most]{tcolorbox} 
\usepackage{pifont}
\usepackage{sistyle}
\usepackage{makecell}
\SIthousandsep{,}
\usepackage{rotating}
\usepackage{hyperref}
\usepackage{colortbl}

\def\ie{i.e.,\xspace}

\def\etc{etc}
\def\eg{e.g.,\xspace}

\newcommand{\oursystem}{\textsf{NeRF$^2$}\xspace}

\newcommand*\diff{\mathop{}\!\mathrm{d}}

\graphicspath{{figs/}{oldfigs/}}
\usepackage{kantlipsum}

\begin{document}
\settopmatter{printacmref=false} 
\renewcommand\footnotetextcopyrightpermission[1]{} 
\title{NeRF$^\textbf{2}$: Neural Radio-Frequency Radiance Fields} 

\settopmatter{authorsperrow=1}
\author{Xiaopeng Zhao, Zhenlin An$\footnotemark[1]$, Qingrui Pan, Lei Yang}
\affiliation{
	\institution{Department of Computing, The Hong Kong Polytechnic University}
	\country{}
}
\email{{zhao,an,pan,young}@tagsys.org}
\renewcommand{\shortauthors}{Xiaopeng Zhao, Zhenlin An, Qingrui Pan, Lei Yang}
\authornote{Zhenlin An and Lei Yang are corresponding authors.}
\def \authors{Xiaopeng Zhao, Zhenlin An, Qingrui Pan, Lei Yang}

\begin{abstract}

Although Maxwell discovered the physical laws of electromagnetic waves 160 years ago, how to precisely model the propagation of an RF signal in an electrically large and complex environment remains a long-standing problem. The difficulty is in the complex interactions between the RF signal and the obstacles (e.g., reflection, diffraction, etc.). Inspired by the great success of using a neural network to describe the optical field in computer vision, we propose a neural radio-frequency radiance field, \oursystem, which represents a continuous volumetric scene function that makes sense of an RF signal's propagation. Particularly, after training with a few signal measurements, \oursystem can tell how/what signal is received at any position when it knows the position of a transmitter. As a physical-layer neural network, \oursystem can take advantage of the learned statistic model plus the physical model of ray tracing to generate a synthetic dataset that meets the training demands of application-layer artificial neural networks (ANNs). Thus, we can boost the performance of ANNs by the proposed turbo-learning, which mixes the true and synthetic datasets to intensify the training. Our experiment results show that turbo-learning can enhance performance with an approximate 50\% increase. We also demonstrate the power of \oursystem in the field of indoor localization and 5G MIMO.
\end{abstract}

%


\acmConference[ACM MobiCom '23]{The 29th Annual International Conference on Mobile Computing and Networking}{October 2--6, 2023}{Madrid, Spain}

\maketitle

\section{Introduction}
\label{section:introduction}

\begin{figure}[t!]
	\centering
	\includegraphics[width=0.75\linewidth]{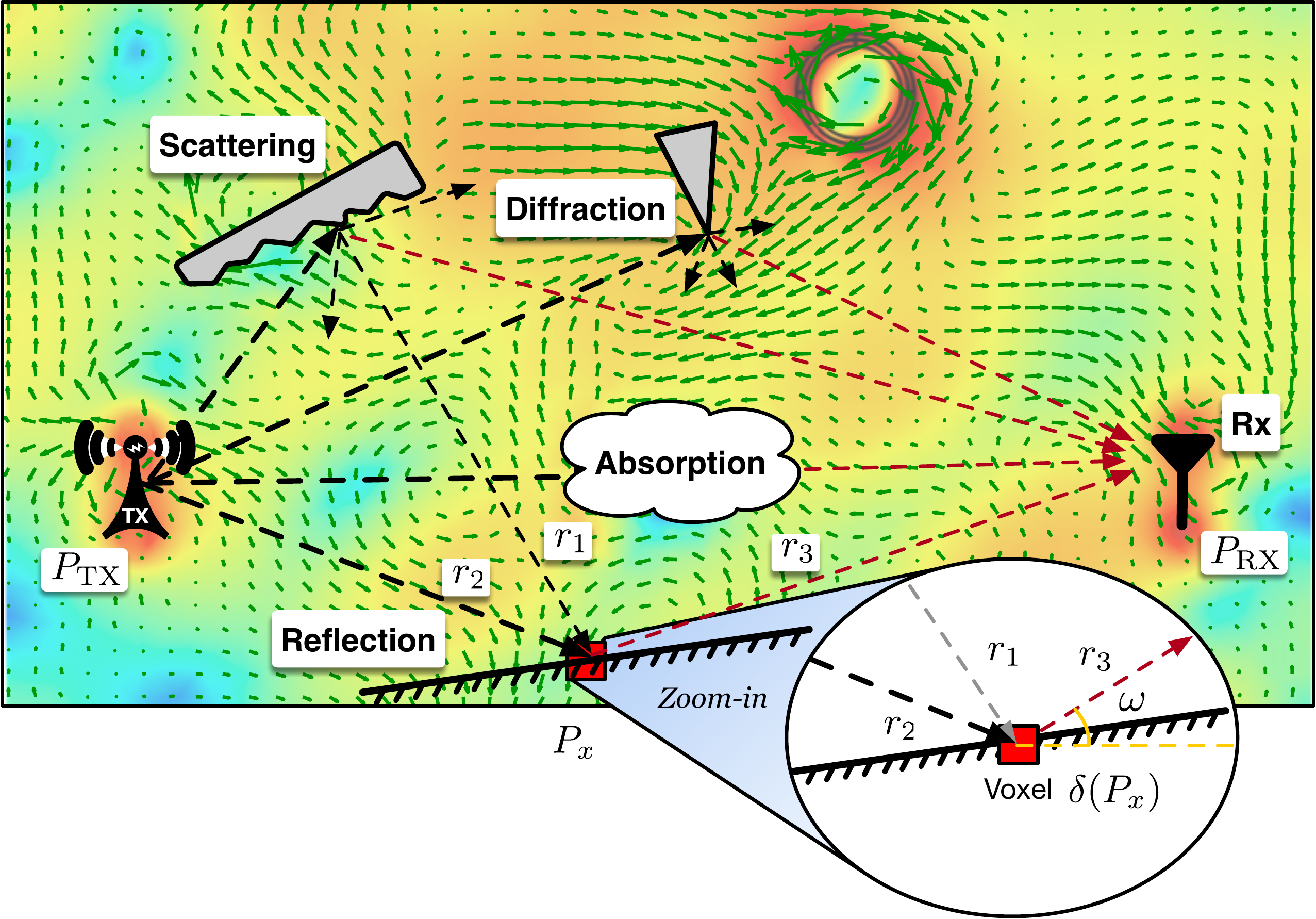}
	\vspace{-0.2cm}
	\caption{Illustration of a radio-frequency radiance field. \textnormal{The ideal distribution of RF radiance is disturbed by the obstacles, which cause the RF signals to be reflected, scattered, diffracted, or absorbed.}}
	\label{fig:propagation}
	\vspace{-0.5cm} 
\end{figure}

In free space, the propagation of an RF signal can be precisely modeled by Maxwell's equation and the Friis equation. However, when objects protrude into the first Fresnel zone defined by the TX and the RX locations, the free-space model fails~\cite{yun2015ray}. As shown in Fig.~\ref{fig:propagation}, the whole radiance field is disturbed by absorption, reflection, diffraction, and/or scattering effects, making electromagnetic ray tracing become extremely complicated. Specifically, (1) reflection occurs on some quasi-specular surfaces (\eg walls, ground, ceiling, \etc.) and follows the law of reflection, \ie the ray is reflected with the angle the same as the incident angle. (2) Diffraction is present at the edges of obstacles and follows the uniform theory of diffraction~\cite{kouyoumjian1974uniform}, \ie a single incident ray upon the edge may create thousands of new rays on Keller cone. (3) The incident ray may also be scattered at small obstacles with rough surfaces where the ray may be reflected toward many angles.

To resolve the above issues, conventional algorithms conduct electromagnetic (EM) ray tracing with a given 3D scene model scanned by LiDAR~\cite{remcom,egea2021opal}. They simulate real EM rays to trace the path that an actual RF signal would take in the real world, which allows better simulation of how the signal interacts with the obstacles in the scene. The tracing quality highly depends on how deep the ray is traced and how realistic the scene model is built.   Unfortunately, RF propagation depends not only on the locations and sizes of obstacles but also on their materials and physical characteristics. Precisely modeling the real-life world using the LiDAR technique is a nearly impossible task in practice.

Recently, Google researchers proposed the neural radiance fields (NeRF) ~\cite{mildenhall2020nerf} to address the ray tracing of light. As one of the important breakthroughs in computer vision,  the NeRF has demonstrated great successes in the view synthesis~\cite{mildenhall2020nerf,tancik2020fourier,barron2021mipnerf}, 3D model rendering~\cite{srinivasan2021nerv,muller2022instant}, and immersive street view  ~\cite{tancik2022block,martin2021nerf}.  A large number of demos can be found at~\cite{nerf-demo}. The basic idea of NeRF is to capture a few images in the scene from different angles as input and then train an MLP (\ie a fully connected class of feedforward artificial neural network) to fit the optical radiance field. The NeRF regards each pixel of an image as a result of one ray tracing, which reflects the feature of the scene-dependent optical radiance field. After training well with a few images, NeRF can exactly predict the result of ray tracing from any other direction and further synthesize an entire image from a given observing direction.

\begin{figure*}[t!]
	\centering
	\includegraphics[width=0.95\linewidth]{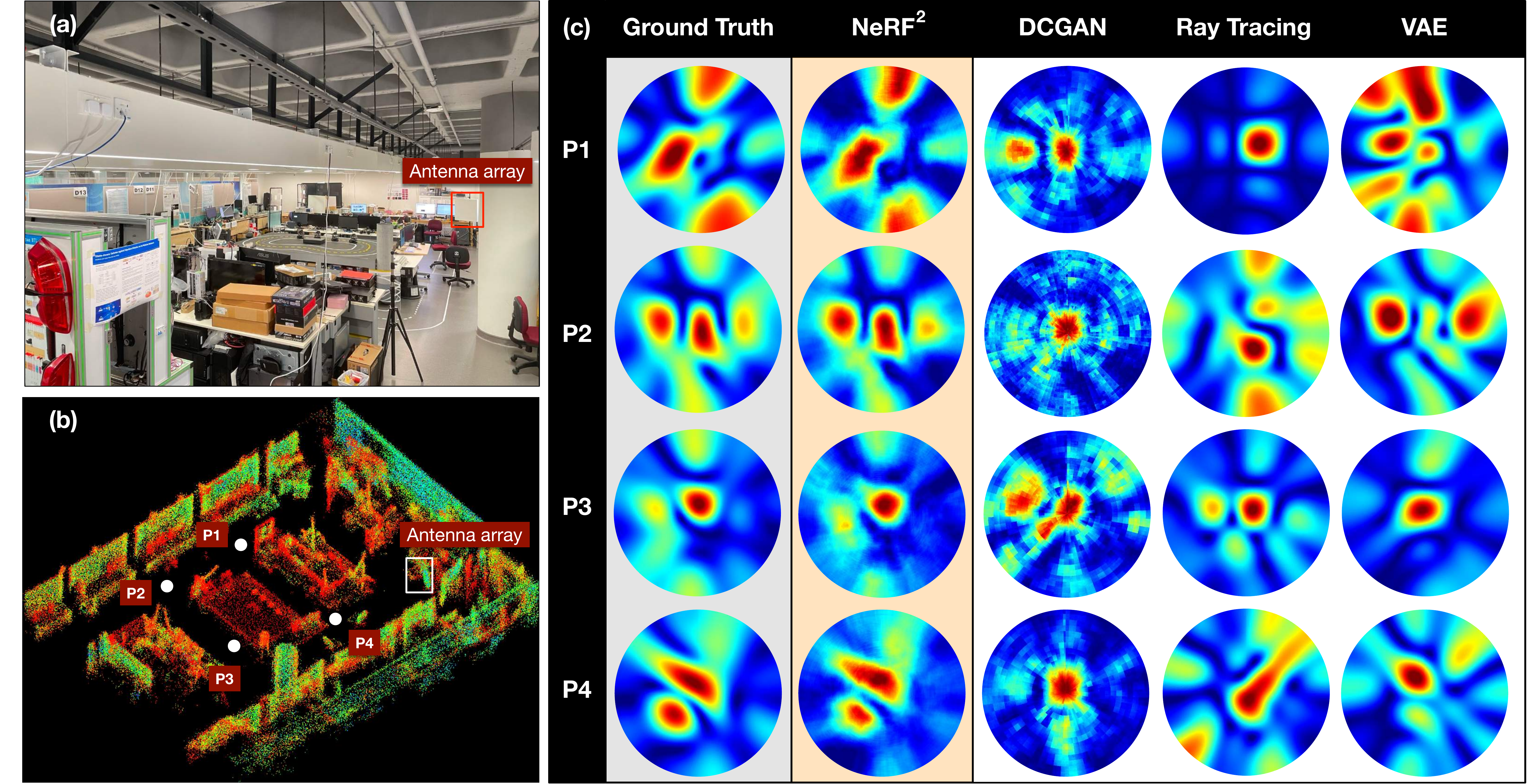}
	\vspace{-0.3cm}
	\caption{Synthesis of spatial spectrums. \textnormal{The spatial spectrums, also known as the multipath profile, show how strong the RF signal is from a particular direction composed of the azimuthal and elevation angles. The formal definition refers to Eqn.~\ref{eqn:spatial-spectrum}. (a) shows the scene, in which the TX may be located at any position, but the RX equipped with a $4\times 4$ antenna array is fixed at a corner; (b) shows the point cloud created by LiDAR, which is only used for the conventional ray-tracing algorithm; (c) compares the synthesis spectrums generated by different algorithms when the TX is located at four different positions. The ground truth is obtained using the antenna array. }}
	\label{fig:predication}
	\vspace{-0.5cm} 
\end{figure*} 

Based on the fact that light is a kind of electromagnetic wave, we propose the \underline{Ne}ural \underline{R}adio-\underline{F}requency \underline{R}adiance \underline{F}ields (\oursystem), which extends the neural radiance fields from optics to electromagnetism. Similarly, \oursystem represents scenes as neural radiance fields by optimizing an underlying continuous volumetric scene function using a sparse set of input signal measurements. Specifically, \oursystem can predict what and how an RF signal is received when the transmitter (TX) is located at a known position. To intuitively understand the capability of \oursystem, we show an example in Fig.~\ref{fig:predication}. Four algorithms are used to synthesize (or predict) the spatial spectrums (\ie multipath profile) that reflect how the RX received the signal from different directions when the TX is located at four different positions. Evidently, the prediction of \oursystem is most similar to the ground truth, which is generated through the true signals received by the antenna array. More examples can be found in our demo video~{\color{red}\url{https://xpengzhao.github.io/NeRF2}}. 

Yet, translating the NeRF to the RF domain requires addressing many challenges. First,  the RF signals operating at UHF or microwave spectrum (\eg 800MHz, 2.4GHz, or 6GHz) are more prone to be reflected, diffracted, and scattered because their frequencies are far lower than the visible light. Second, only the amplitude (\ie light strength) is considered in the optical NeRF. The phase of light is neglected because it repeats every 600--800 nm propagation. By contrast, the phase cannot be disregarded anymore in cm- or mm-wavelength RF signals for its crucial role in constructive or destructive superimposing owing to the multipath effects. Third, the measurement of visible light is taken by using a million-pixel camera, but an RF RX is usually equipped with either a single antenna or a small antenna array because of the size limitation (\ie the size of an antenna is proportional to the wavelength). To address these issues, we first update the physical tracing model to fit the characteristic of RF signals. We then introduce the phase apart from the amplitude to set up a complex-valued MLP for \oursystem. We finally propose two training approaches for the single-antenna and array-antenna receivers.

As a physical-layer neural network, \oursystem can promote the performance of many key RF applications, such as indoor localization, channel estimation, wireless power transmission, 5G base station deployment, wireless sensing, and so on. To meet the various application-layer demands, we propose \emph{turbo-learning}, which takes advantage of the physical nature of \oursystem to generate a vast number of synthetic datasets in accordance with the physical model. This synthetic dataset is mixed with the true dataset together to intensify the training of application-layer artificial neural networks (ANNs). Turbo-learning not only allows ANNs to collect fewer training datasets but also promises a high-level learning accuracy. 
 
\textbf{Summary of results}. We use a $4\times 4$ antenna array as the RX to predict the spatial spectrums (\ie multipath profile) in the micro benchmark. The results show that the median similarity of the spatial spectrums generated by \oursystem and the ground truth is up to 82\%, which is far higher than other synthetic algorithms. We also use light-of-sight (LOS) AoA estimation as an application to quantify the benefit of turbo-learning. The experiment results show that the accuracy can be raised by 47.9\% using only a 10\% true training dataset. Our large-scale experiments in which we collect RF signals at 530K positions in 14 scenes further verify the great power of turbo-learning. Overall, the average AoA accuracy is improved by 49.5\% across 14 scenes. 

\textbf{Field Study}. We present two field studies to demonstrate how the \oursystem benefits two classical applications: (1) \ul{BLE Localization}. Pinpointing an RF device indoors is challenging~\cite{yang2014tagoram,ma2017minding,xie2019md,xie2018swan,adib20133d,adib2014multi,zhao2018rf,zhao2018through,ma2014accurate,hui2019radio,haniz2017novel,youssef2005horus,sen2012you,yang2012locating,liu2012push,wang2012no,ni2004landmarc,wang2013dude,pan2008transfer}, particularly when the line-of-sight propagation is blocked. Similar to the problem of ray tracing in graphics, localization accuracy can be improved greatly if the propagation of RF signals is deeply traced using \oursystem. Our experiment results show that the \oursystem enabled turbo-learning can reduce the median error by 50\% and the standard variance by 40\%. (2) \ul{5G MIMO}. Massive MIMO (\ie 5G network) heavily relies on accurate channel state information (CSI) for beamforming, \ie the base stations must know the downlink wireless channel from their antennas to every client devices~\cite{liu2021fire}. To this end, the client devices should transmit the channel estimation results back to the base station and thus cause huge overheads. The demand channel estimation can be exactly met by the \oursystem in that it can predict the CSI at any position derived from the learned radiance fields. Our experiment results show that \oursystem is  5.97 dB better SNR and 4.32 dB better SINR than the state-of-the-art work in terms of channel prediction and MU-MIMO performance. 

\textbf{Contribution}. Our contributions are summarized as follows. 
\begin{itemize}[leftmargin=*]

\item We translate the NeRF from optics to the RF domain. Specifically, (1) We update the neural networks for complex-valued input parameters; (2) we replace the light propagation model with the Friis equation; (3) we invented the single-antenna and multiple-antenna-based electromagnetic ray-tracing approaches.

\item We propose the \oursystem enabled turbo-learning. The benefit of turbo-learning is not only limited to the performance enhancement but also, more importantly, addresses the pain point of deep learning --\ significant reduction of the quantity of training dataset and the corresponding workload on collecting dataset.

\item We conduct real-life field studies in terms of RFID, BLE, and 5G systems. The proposed turbo-learning is evaluated on indoor localization and FDD massive MIMO channel prediction.
\end{itemize}

\section{N\lowercase{e}RF$^\textbf{2}$ Design}
\label{section:design}

Following a common practice of NeRF, we make the following similar assumptions:  (1) The receivers (\eg 5G base station, Bluetooth station, and RFID reader) are located at known positions, whereas the transmitter (\eg smartphones, iBeacon, and RFID tags) are movable within a limited range. (2) Major obstacles (\eg buildings, walls, and furniture) in each scene remain unchanged.  (3) The moving obstacles may introduce temporary minor perturbance on the radiance field, which can be smoothed through the upper-layer filtering algorithm (\eg Kalman filter), so their influence is not considered. 

At the heart of \oursystem is the two key components: the neural radiance network and the ray tracing algorithm:
\begin{itemize}[leftmargin=*]	
  \setlength{\parskip}{0pt}
  \setlength{\itemsep}{0pt plus 1pt}
  \item \textbf{Neural Radiance Network}:  This network is used to represent the scene and the radiance field using two MLPs. It can predict how RF signals are distributed in the scene.
  \item \textbf{Ray Tracing}: Given the RF distribution, we must trace the signals transmitted from all potential directions to know what signal is received at the RX. 
\end{itemize}
In this section, we elaborate on the details of the above two components and present the training approaches. Finally, we present \oursystem enabled turbo-learning.

\subsection{Neural Radiance Network}

To model the radiance field,  we discretize the scene of interest into a finite number of small 3D voxels in the space. The Huygens--Fresnel principle suggests that when the original RF signal arrives at it from all possible paths, a voxel can be considered as a new radiance source retransmitting the RF signal. To better understand this principle, we show an example voxel in the zoom-in of Fig.~\ref{fig:propagation}. Let the subscript $x$ denote an arbitrary voxel in the scene. The voxel $x$ at position $P_x$ receives the RF signal from two paths $r_1$ and $r_2$, and it becomes a new TX that retransmits the RF signal along the path $r_3$ to the RX. In our model, each voxel is described with three properties, the position $P_x=(X,Y,Z)$,  the attenuation $\mathbf{\delta}(P_x) = \Delta a (P_x) e^{\mathbf{J}\Delta \theta (P_x)}$ and the retransmitted RF signal $S(P_x)$. The $\delta (P_x)$ is a material-dependent variable, which indicates that the amplitude is degraded by $\Delta a (P_x) = |\delta (P_x)|$ and the phase is rotated by $\Delta \theta (P_x)=\angle \delta (P_x)$ if an RF signal passes through the voxel at $P_x$. As a new RF transmitter, the voxel at position $P_x$ retransmits a new complex-valued signal $S_x$,  \ie $S(P_x)=a(P_x) e^{\mathbf{J}\theta(P_x)}$ where $\theta(P_x)$ and $a(P_x)$ are the initial phase and the initial amplitude.  The voxel cannot be simply modeled as an omnidirectional radiance source. Instead, it may radiate EM waves unevenly in angles. To address this issue,  we introduce another variable called measuring direction $\mathbf{\omega}=(\alpha,\beta)$ where $\alpha$ and $\beta$ are the azimuthal and elevation angles. As shown in the zoom-in of Fig.~\ref{fig:propagation}, the voxel is located in the direction $\omega$ relative to the RX position. 

 \oursystem aims to predict the RF signal $S_x$ retransmitted from the voxel at $P_x$ toward the direction $\omega$  when given the  TX's position $P_\text{TX}$. To do so, we take advantage of the neural network to fit the radiance field. Formally, the radiance field $\mathbf{F}$ is represented as follows:
\begin{equation}\label{eqn:F}
	\mathbf{F}_{\Theta}:  (P_\text{TX},  P_x, \omega) \rightarrow \bigg(\delta (P_x), S(P_x, \omega)\bigg) 
\end{equation}
where $\Theta$ indicates the learnable neural network weights. Unlike the visual NeRF that assumes the ambient light remains unchanged, we introduce the position of the TX as an additional input because our transmitters (\eg smartphones or IoT devices) are moveable. In this way, we can create a dataset for a scene by placing the TX at different and sufficient positions. The neural network contains two outputs. One is the attenuation $\delta(P_x)$ of the voxel at $P_x$, which is highly related to the voxel's physical characteristics. Another is the RF signal $S(P_x, \omega)$ retransmitted from the voxel at $P_x$ toward the direction $\omega$. Thus, the neural network represents not only the scene but also the RF distribution.

\begin{figure}[!t]
	\centering
	\includegraphics[width=0.95\linewidth]{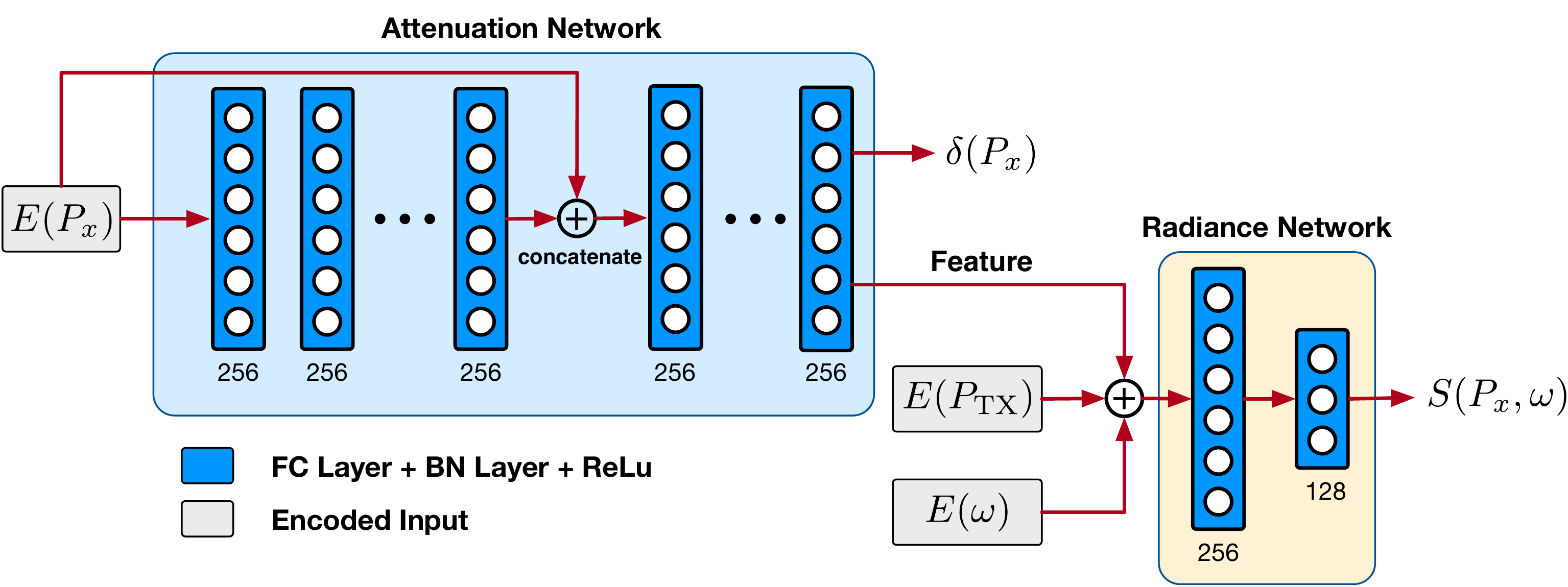}
	\vspace{-0.3cm} 
	\caption{Architecture of the neural network. \textnormal{\oursystem consists of two MLPs, the attenuation network, and the radiance network. The attenuation network can predict the attenuation $\delta$ of any voxel. Given the TX position and a measuring direction,  the radiance network can predict the signal transmitting from an arbitrary voxel.}}
	\label{fig:nn-structure}
	\vspace{-0.5cm}
\end{figure}

\begin{figure*}[t!]
	\centering
	\includegraphics[width=0.9\linewidth]{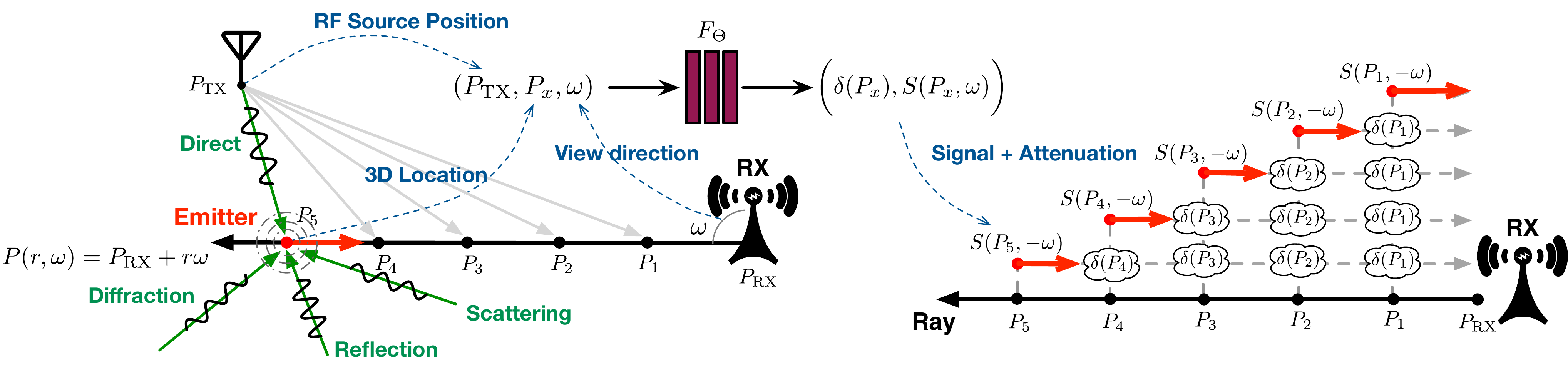}
	\vspace{-0.5cm}
	\caption{Electromagnetic ray tracing. \textnormal{There are five voxels at $P_1 - P_5$ on the ray. Each voxel becomes a new transmitter that emits the signal along the ray to the RX. Their signals are attenuated by the other voxels between the new transmitters and the RX. }}
	\label{fig:architecture}
	\vspace{-0.5cm} 
\end{figure*}

\textbf{Network Architecture}. To build the neural network, we adopt two MLPs: the attenuation network and the radiance network, as shown in Fig.~\ref{fig:nn-structure}. The attenuation property is highly related to the materials of the voxel and independent of the incoming signals, so we separate the attenuation network to predict the attenuation $\delta(P_x)$ as a function of the position $P_x$. The attenuation network is composed of eight fully connected layers (using ReLU activations and 256 channels per layer) and outputs $\delta(P_x)$ and a 256-dimensional feature vector. This feature vector is then concatenated with the RX direction $\omega$ related to $P_x$, and the TX position $P_\text{TX}$. The combination is passed to the radiance network, another two fully connected layers (using a ReLU activation and including 256 and 128 channels), which outputs the direction-dependent RF signal $S(P_x, \omega)$, which is retransmitted from the voxel along the direction $\omega$. The network architecture is similar to the optical NeRF but differs in two aspects. First, the visual NeRF assumes that the location of TX (\ie light source) remains unchanged, whereas our TX is moveable. Second, our two networks are complex-valued, considering both magnitude and phase. 

\textbf{Discussion}. The radiance field is only relevant to the scene, including the obstacles and the position of TX, but irrelevant to the position of RX. One may be concerned about how to deal with the multiple reflections of the RF signal. The trick of \oursystem is that each voxel is considered as a new transmitter that ``retransmit'' a combined signal received from all possible paths. Such a model simplifies the subsequent calculation of ray tracing.  

\subsection{Electromagnetic Ray Tracing}

To train the \oursystem, a naive approach is to probe the RF signals at a vast number of RX positions. Evidently, this approach is unscalable in practice. The visual NeRF views each image of the scene as a result of ray marching~\footnote{Ray tracing and ray marching are two rendering techniques in computer graphics. Ray tracing computes the resulting color by tracing rays and accounting for object interactions, while ray marching estimates the color and opacity of the scene by evaluating a function along a ray.}, where each pixel reflects the intensity of the light propagated from a particular direction due to the pinhole model of cameras. Similarly, the signal received at the RX is a result of electromagnetic ray tracing, where the signal is a combination of signals transmitted from all possible directions. Next, we introduce how we trace the signal from a particular direction.

The propagation of an RF signal $S$ from a transmitter (TX) to a receiver (RX) conforms to the Friis equation as follows:
\begin{equation}\small
\label{eqn:friis}
		R = H_{\text{TX}\rightarrow \text{RX}}S = a_{\text{TX}\rightarrow \text{RX}} e^{\mathbf{J} \theta_{\text{TX}\rightarrow \text{RX}}}S
\end{equation}
where $R$ is the received signal, $H_{\text{TX}\rightarrow \text{RX}}$ is the channel attenuation. Particularly, $ a_{\text{TX}\rightarrow \text{RX}}$ and $\theta_{\text{TX}\rightarrow \text{RX}}$ are the amplitude degradation and the phase rotation caused by the distance from the TX to the RX. Mathematically, a direction $\omega$ related to the RX can be modeled as a ray, which starts from the RX and directs toward $\omega$. The points on this ray are correspondingly described as follows:\begin{equation}\small
	P( r, \omega) = P_\text{RX}+r\omega
\end{equation}
where $r$ is the radial distance from the RX to the point on the ray. Note that $P_\text{RX}= P(0,\omega)$. The purpose of ray tracing is to accumulate the RF signals emitted from all voxels on this ray. Namely, the received signal at the RX from the direction $\omega$ can be expressed as:
\begin{equation}\label{eqn:r}\scriptsize
	R(\omega) = \int_0^{D} H_{P( r, \omega)\rightarrow P_\text{RX}} S\bigg(P(r, \omega), -\omega\bigg) \diff r
\end{equation}
In the above equation, $S(P(r, \omega), -\omega)$ represents the signal transmitted from the voxel at $P(r, \omega)$ to the RX at $P_\text{RX}$. Its transmission direction is opposite to the ray's direction, so we take the negative of $\omega$ in the equation. $D$ is the maximal distance across the scene. The above equation suggests that the final signal received by the RX from the direction $\omega$ is the accumulation of the RF signals transmitted from all voxels on the ray, \ie from $P(0, \omega)$ to $P(D, \omega)$. The $H_{P(r,\omega)\rightarrow P_\text{RX}} $ is the attenuation of the signal propagated from the point $P(r, \omega)$ to the RX. It is defined as follows:
\begin{equation}\footnotesize
\begin{aligned}
	H_{P(r, \omega)\rightarrow P_\text{RX}} & = \prod_{\tilde{r}=0}^r \delta\big(P(\tilde{r}, \omega)\big)  \\
	&=\bigg(\prod_{\tilde{r}=0}^r \Delta a_{P(\tilde{r}, \omega)} e^{\mathbf{J}\Delta \theta_{P(\tilde{r}, \omega)}}\bigg)\raisetag{3\baselineskip}
\end{aligned}
\end{equation}  
The above equation means that the total attenuation equals the product of all attenuations caused by the voxels between the voxels at $P(r,\omega)$ and at $P(0,\omega)$, \ie $0 \leq \tilde{r} \leq r$. To facilitate the calculation, we transform the above equation to an equivalent log-scale form as follows:
\begin{equation}\footnotesize
	\begin{split}\label{eqn:h}
	H_{P(r, \omega)\rightarrow P_\text{RX}} & = \text{exp}\Bigg( \ln \bigg(\prod_{\tilde{r}=0}^r \delta \big(P\big(\tilde{r}, \omega)\big)\bigg)\Bigg) = \text{exp}\Bigg(\int_{0}^{r} \ln\bigg(\delta\big(P(\tilde{r}, \omega)\big)\bigg) \diff \tilde{r}\Bigg)\\
	&=	\exp  \left(  \underbrace{ \int_{0}^{r} \hat{\delta}\bigg({P(\tilde{r}, \omega)\bigg)}\diff \tilde{r}}_{\text{Sum of attenuations}} \right) \raisetag{3\baselineskip}
\end{split}
\end{equation}
where $\hat{\delta}(\cdot)$ denotes the log-scale attenuation of $\delta(\cdot)$, which is defined as follows:
\begin{equation}\footnotesize
\begin{aligned}
	\hat{\delta} (P(\tilde{r}, \omega)) &=\ln \delta (P(\tilde{r}, \omega))
\end{aligned}
\end{equation}
The log-scale form makes the product become a sum of all attenuations between two voxels, which greatly facilitates the calculation. Substituting Eqn.~\ref{eqn:h} into Eqn.~\ref{eqn:r}, the signal coming from the direction $\omega$ is given by
\begin{equation}\scriptsize
	R(\omega) =  \int_0^{D} \underbrace{ \exp  \left(\int_{0}^{r} \hat{\delta}\big({P\big(\tilde{r}, \omega)\big)}\diff \tilde{r}\ \right) 	}_{\text{Attenuation Network}}  \overbrace{S(P(r,\omega), -\omega)}^{\text{Radiance Network}}\diff r     
\end{equation}
where the terms engaged in the previous part are predicted by the attenuation network, and the terms engaged in the last part are predicted by the radiance network. Briefly, the result of ray tracing along a direction is to aggregate the signals retransmitted from the voxels on this ray, each of which is regarded as a new source. Meanwhile, each transmission from a voxel must be attenuated by other voxels between the current voxel and the RX. Suppose there are $N$ voxels on the ray,  the ray tracing will take $\mathcal{O}(N^2)$ aggregations. 

To visually understand the ray tracing algorithm, we show an example in Fig.~\ref{fig:architecture}. Assuming the horizontal ray is from the RX to the left (\ie $\omega=180^\circ$). On the ray, there are five voxels at $P_1$, $P_2$, $P_3$, $P_4$, and $P_5$, all of which are considered as new transmitters regardless of how these voxels are lighted up. As a result, the signal received by RX along the opposite direction of the ray  (\ie $-\omega$) is a combination of the five signals retransmitted from these five voxels. Particularly, the signal $S_5$ retransmitted from the voxel at $P_5$ is attenuated by the voxels at $P_4$, $P_3$, $P_2$, and $P_1$ in sequence. The accumulated attenuation equals $(\hat{\delta}_{P_1}+\hat{\delta}_{P_2}+\hat{\delta}_{P_3} +\hat{\delta}_{P_4})$. Similarly, the signals retransmitted from the voxels at $P_1$, $P_2$, $P_3$, and $P_4$ are attenuated by $0$, $\hat{\delta}_{P_1}$, $\hat{\delta}_{P_1}+\hat{\delta}_{P_2}$, and $\hat{\delta}_{P_1}+\hat{\delta}_{P_2}+\hat{\delta}_{P_3}$, respectively.

\textbf{Summary}. The \oursystem does not completely depend on the neural network but combines the physical model and the statistic model. Specifically, ray tracing takes a well-known physical model of signal propagation, meanwhile, deep learning offers a statistical model of the complicated interactions between the RF signal and the surrounding obstacles.  

\subsection{Network Training}

The previous describes the ray tracing algorithm, by which we can use the \oursystem to predict the signal received by the RX from a particular direction. Regarding which type of antenna is equipped at the RX, we introduce two types of training approaches.  

\subsubsection{Case I: Single-Antenna RX Model}
\label{section:single-antenna}

We consider a simplified case where the RX is equipped with a single omnidirectional or a single directional antenna. Evidently, a single antenna has no discernibility in directions. Thus, the eventually received signal by the RX is a combination of the signals from all potential directions as follows:
\begin{equation}\scriptsize\label{eqn:single-antenan-model}
\begin{split}
	R&=\int_{\Omega} \sqrt{G_\text{RX}(\omega)}R(\omega)\diff \omega \\
	&= \int_\Omega \int_0^{D} \exp  \left(\int_{0}^{r} \hat{\delta}({P\big(\tilde{r}, \omega)\big)}\diff \tilde{r}\ \right) S(P(r,\omega), -\omega)\diff \tilde{r}     
\end{split} 
\end{equation}
where $G_\text{RX}(\omega)$ indicates the antenna directivity (\ie the gain that the antenna provides in each direction), and $\Omega$ denotes the directions that the antenna can cover. Let $R$ and $\widetilde{R}$ denote the predicted signal by \oursystem with the ray tracing and the true received signal, respectively. We then can use the following loss function to train \oursystem:
\begin{equation}\small
	\mathcal{L} = |R- \widetilde{R}|^2
\end{equation}
The loss function aims to reduce the gap between the true signal and the predicted one.

\subsubsection{Case II: Multi-Antenna RX Model}
\label{section:muti-antenna}

\begin{figure}[t!]
  \centering
    \subfigure[3D Spatial Spectrum]{\label{fig:spatial-3d}
    \includegraphics[height=3cm]{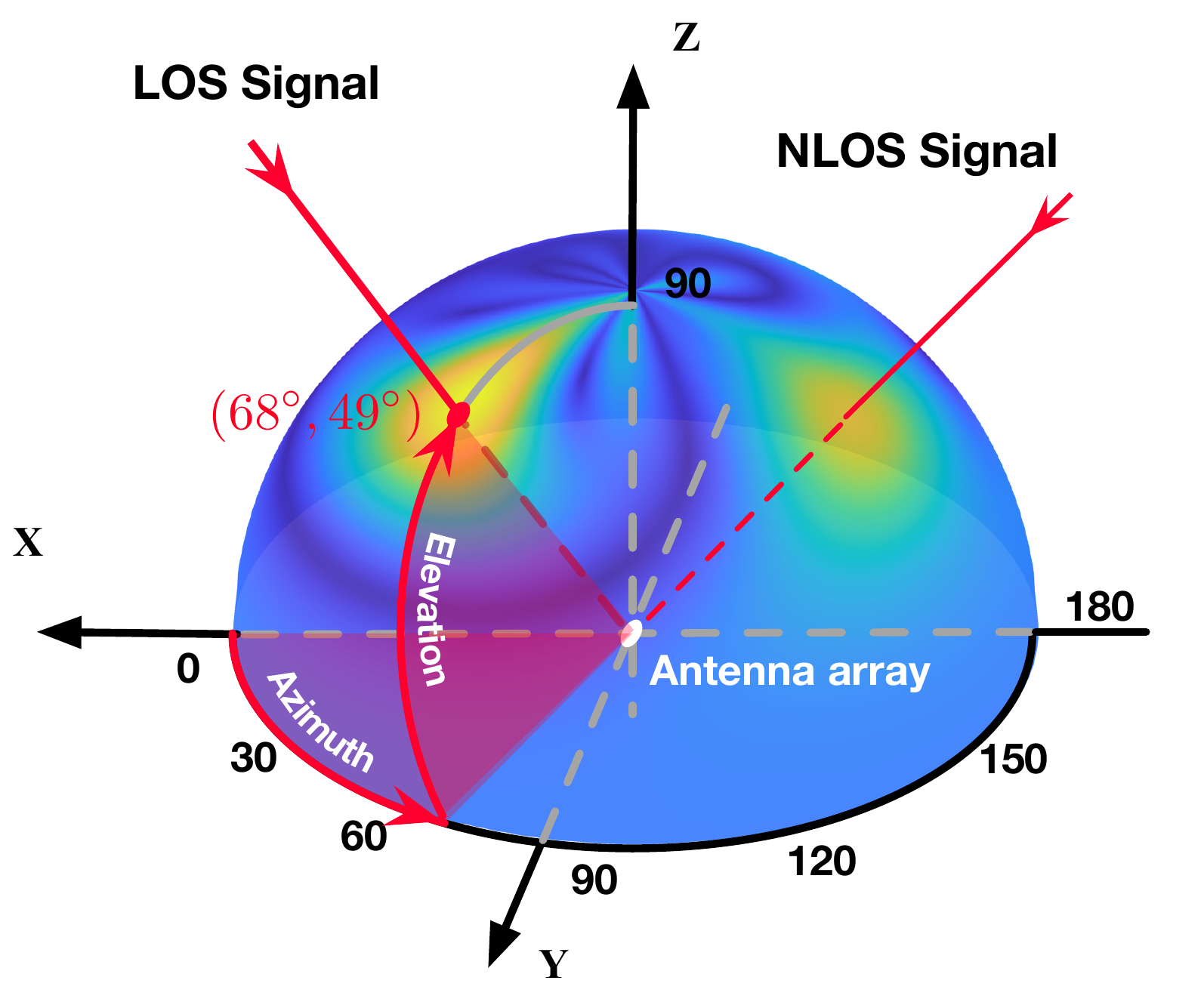} 
   }%
   \subfigure[2D Spatial Spectrum]{\label{fig:spatial-2d}
    \includegraphics[height=3cm]{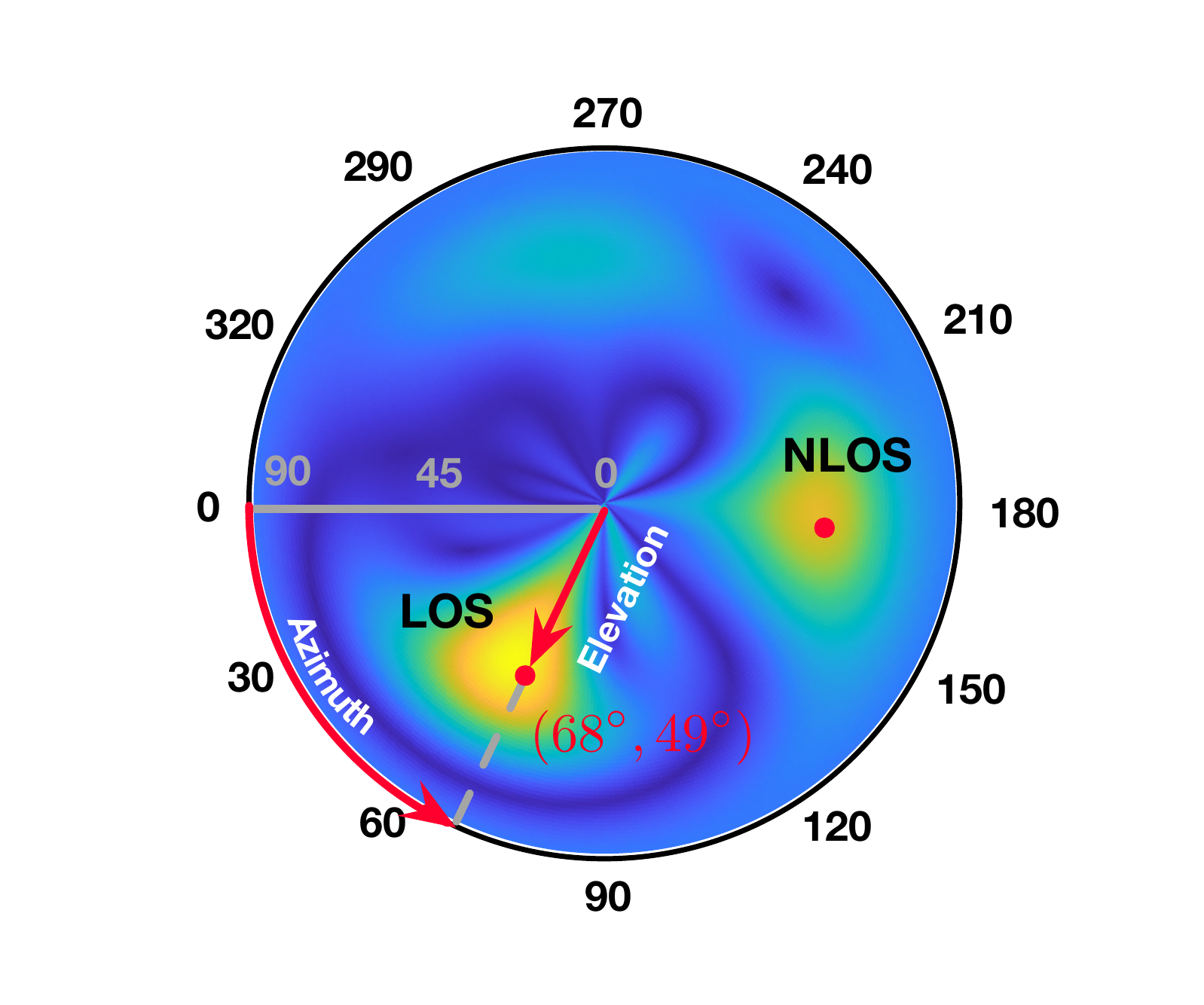}  
   }
   \vspace{-0.5cm}
   \caption{Illustration of spatial spectrum}
   \label{fig:spatial-spectrum}
   \vspace{-0.5cm}
\end{figure}

Next, we consider the second case where the RX is equipped with a phased antenna array, which can form a very narrow beam and steer it to receive signals from a particular direction~\cite{stoica2005spectral}. The RX can then discriminate the signal in directions. 
Suppose the antenna array is equipped with $K\times K$ elements uniformly. Choosing the element $A_{1,1}$ as a reference, we can compute the following relative power of projecting the received signal into the direction of $\omega = (\alpha, \beta)$: 
\begin{equation}\scriptsize
\label{eqn:relative-power}
	\Psi(\omega) = \frac{1}{\left(K^2-1\right)}\left| \sum_{i=1}^{K}\sum_{j=1}^{K} w_{i,j}(\omega)\cdot e^{\rm{J} \Delta \widetilde{\theta}_{i,j}} \right|
\end{equation}
\noindent where $w_{i,j}(\omega)=e^{\rm{J} -\Delta \theta_{i,j}}$ is the complex weight for steering a beam to a certain angle of $(\alpha,\beta)$. In the above, $\Delta\widetilde{\theta}_{i,j}$ is the phase difference computed by using the received signals at $A_{i,j}$ and $A_{1,1}$, whereas $\Delta{\theta}$ is their theoretical phase difference~\cite{an2020general}. The sum aggregates the relative power across the $(K^2-1)$ pairs of elements, \ie $(A_{1,2}, A_{1,1}), (A_{1,3}, A_{1,1}), \cdots$.  When $\Delta\widetilde{\theta}_{i,j}$ aligns with $\Delta{\theta}_{i,j}$, \ie the signal comes from the direction of $(\alpha, \beta)$, the normalized relative power $\Psi(\alpha, \beta)$ should achieve the maximum. A heatmap can then be generated to show the relative power at $N$ possible directions that the received RF signal might come from. We call such a 2D heatmap \emph{spatial spectrum}, denoted by $\Psi$.
The $N$ is a custom parameter depending on the angle resolution. If the one-degree resolution is accepted, $N=360\times 90$, and the spatial spectrum is defined as follows:
\begin{equation}\scriptsize\label{eqn:spatial-spectrum}
	\Psi = 
\begin{pmatrix}
	\Psi(0^\circ,0^\circ) &\Psi(1^\circ,0^\circ)& \cdots & \Psi(360^\circ, 0^\circ) \\
    \Psi(0^\circ,1^\circ)& \Psi(1^\circ,1^\circ)& \cdots & \Psi(360^\circ, 1^\circ) \\
     \vdots & \vdots &\vdots & \vdots \\
    \Psi(0^\circ,90^\circ) & \Psi(1^\circ,90^\circ) & \cdots & \Psi(360^\circ, 90^\circ)
\end{pmatrix}
\end{equation}
Sometimes, the spatial spectrum is also called multipath profile~\cite{wang2013dude} because it reflects how the signal comes from multiple directions. 
Fig.~\ref{fig:spatial-3d} shows the spatial spectrum in 3D where all directions are uniformly distributed; Fig.~\ref{fig:spatial-2d} shows the 2D spectrum by projecting the 3D onto the X-Y plane, in which the radial distance represents $\cos(\beta)$ so the elevation angle distributes non-uniformly.

It is spontaneous for \oursystem to predict the  power of the signal coming from a particular direction and generate a predicted spatial spectrum $\Psi'$ as follows:
\begin{equation}\small
	\Psi'(\omega) =  	|R(\omega)|^2
\end{equation}
The relative power is directly proportional to the true power computed above. Even though a constant offset may exist between them, it does not affect the training of the network by using the following loss function:
\begin{equation}\label{eqn:case2}\scriptsize
	\mathcal{L} = \sum_{\omega\in \Omega}|\Psi(\omega)-\Psi'(\omega)|^2
\end{equation}
The training aims to reduce the difference in power of the signal received from all possible directions.

\subsection{Turbo-Learning}
\begin{figure}[!t]
	\centering
	\includegraphics[width=0.85\linewidth]{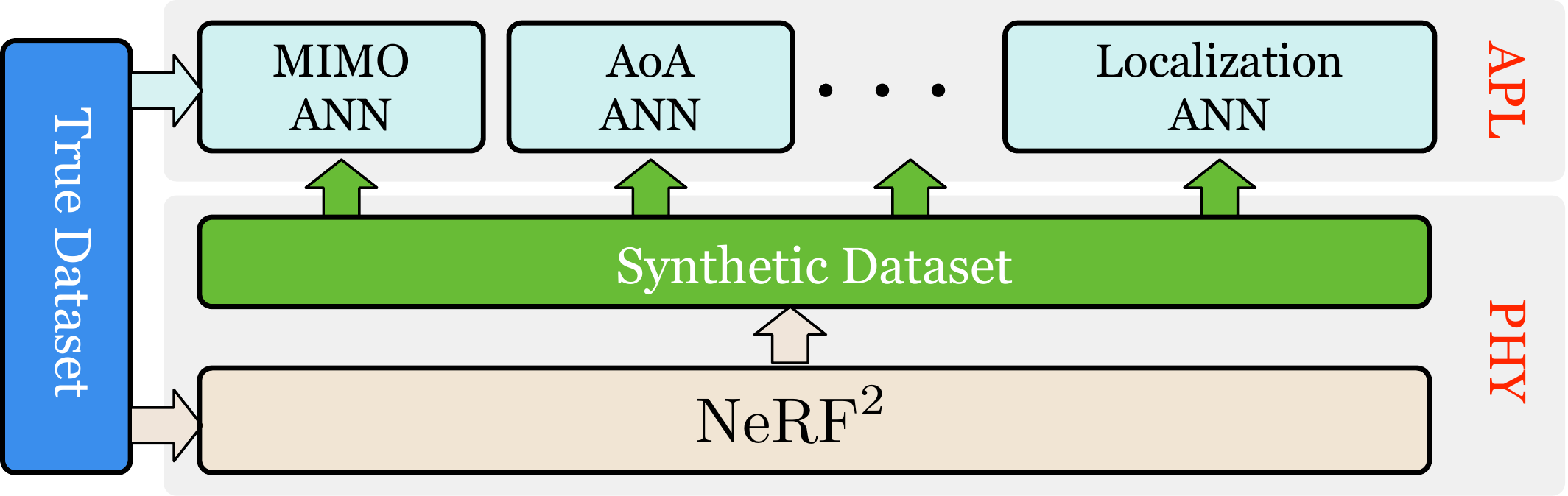}
	\vspace{-0.4cm} 
	\caption{Illustration of turbo-learning}
	\label{fig:turbo-learning}
	\vspace{-0.5cm}
\end{figure}

As a physical-layer neural network, \oursystem describes the distribution of the radiance field. It cannot directly meet the application-layer demands, such as predicting the location of a receiver or beamforming parameters. Usually, extra neural networks are set up to address the specific application demand  (\eg  MIMO ANN, AoA ANN, localization ANN, \etc.).  Instead, we employ \oursystem as a reinforcer to boost the performance of the application-layer ANNs. Fig.~\ref{fig:turbo-learning} illustrates this basic idea. First, we train \oursystem with the true training dataset. Second, \oursystem generates a vast number of synthetic dataset which meets the demand of the application-layer ANNs. Finally, we mix the true dataset and the synthetic dataset together to train the upper-layer ANNs. We call this training approach \emph{turbo-learning}, \ie applying more additional synthetic data to intensify the learning. Turbo-learning is also termed data augmentation in the field of data science. In the following sections, we will elaborate on turbo-learning case by case. 
\begin{figure*}[t!]
  \centering
     \subfigure[10\% TS + 90\% SS]{\label{fig:aoa-accuracy-10}
    \includegraphics[height=3cm]{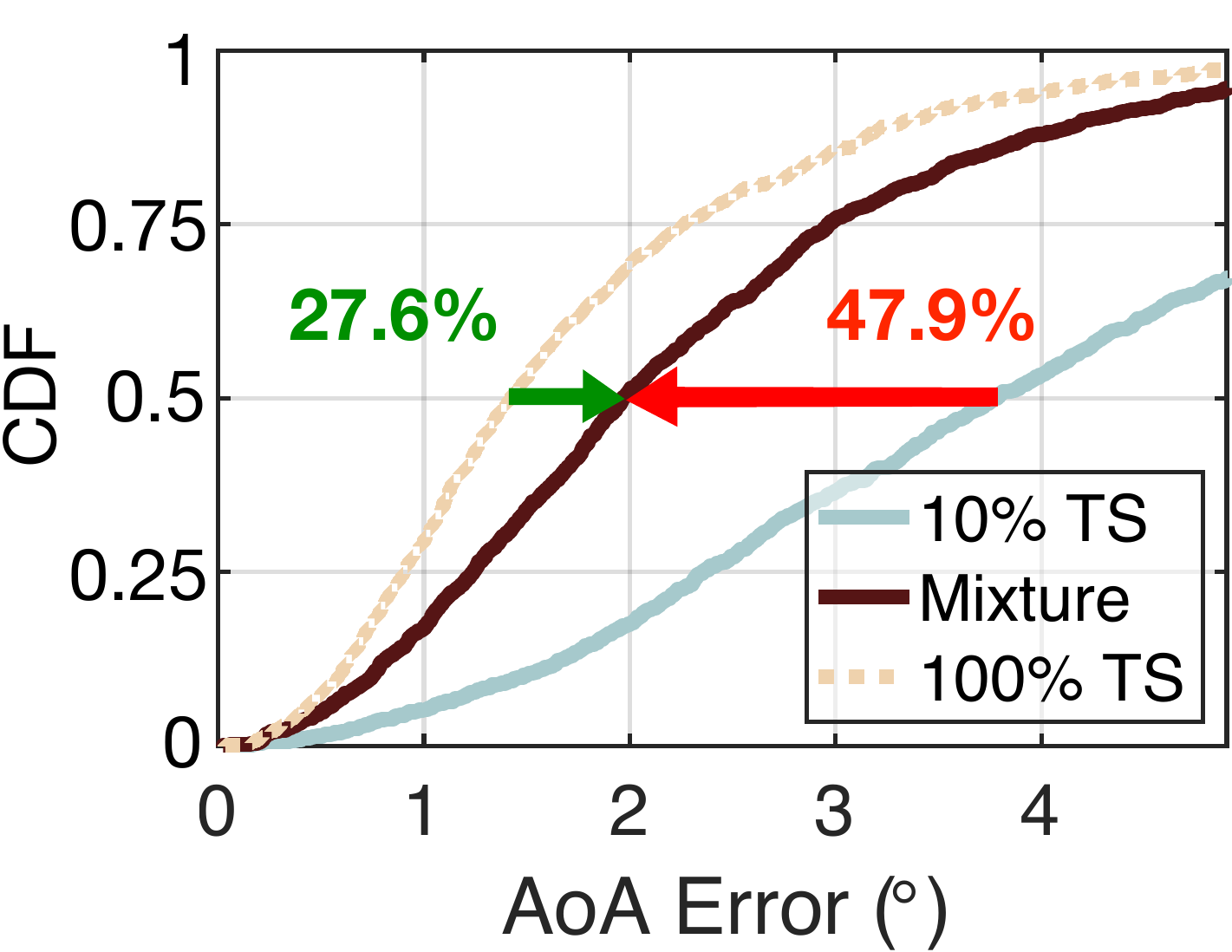}  
   }%
   \subfigure[30\% TS + 70\% SS]{\label{fig:exp-s23-30}
    \includegraphics[height=3cm]{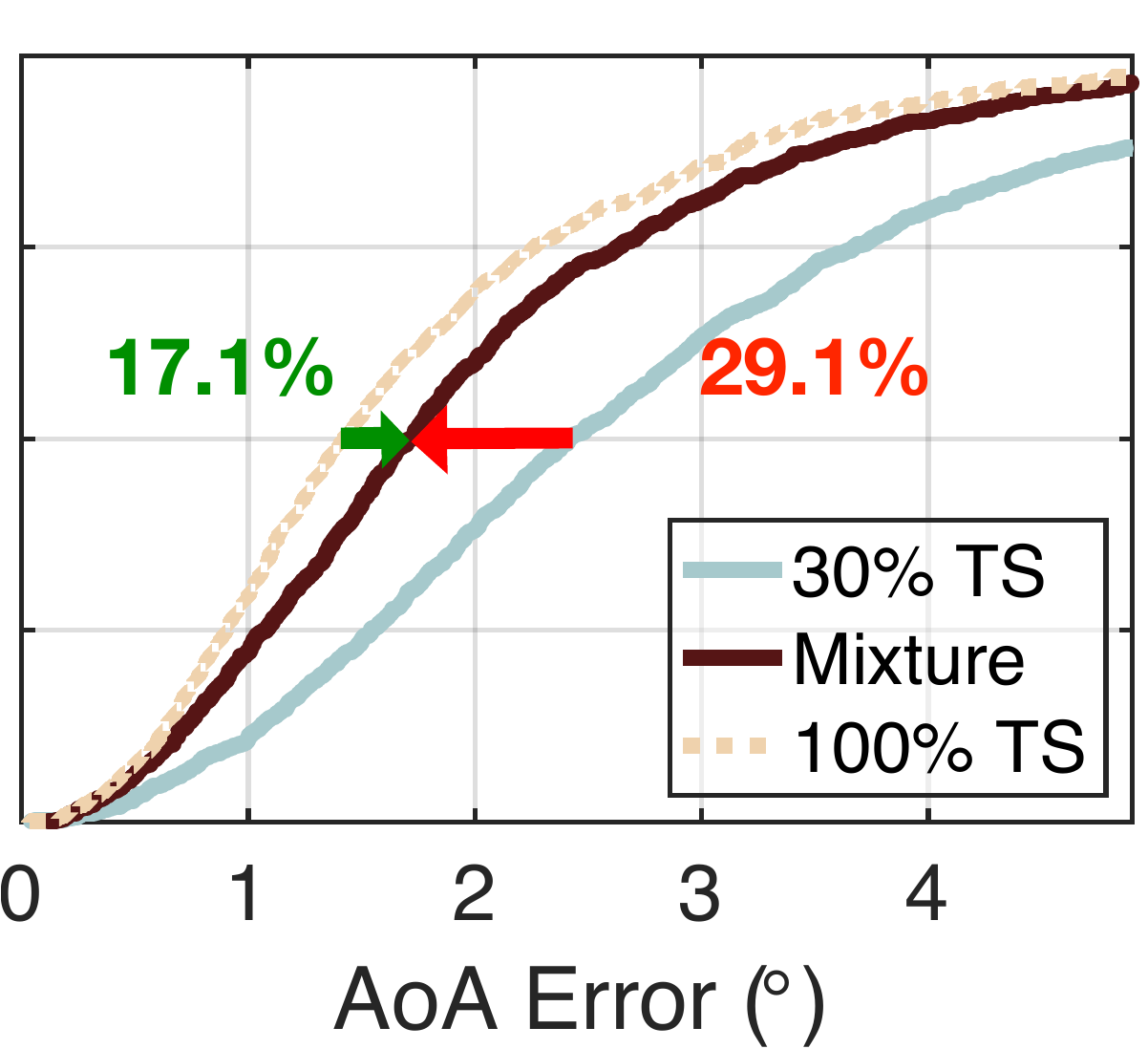}    
   }%
    \subfigure[50\% TS + 50\% SS]{\label{fig:exp-s23-50}
    \includegraphics[height=3cm]{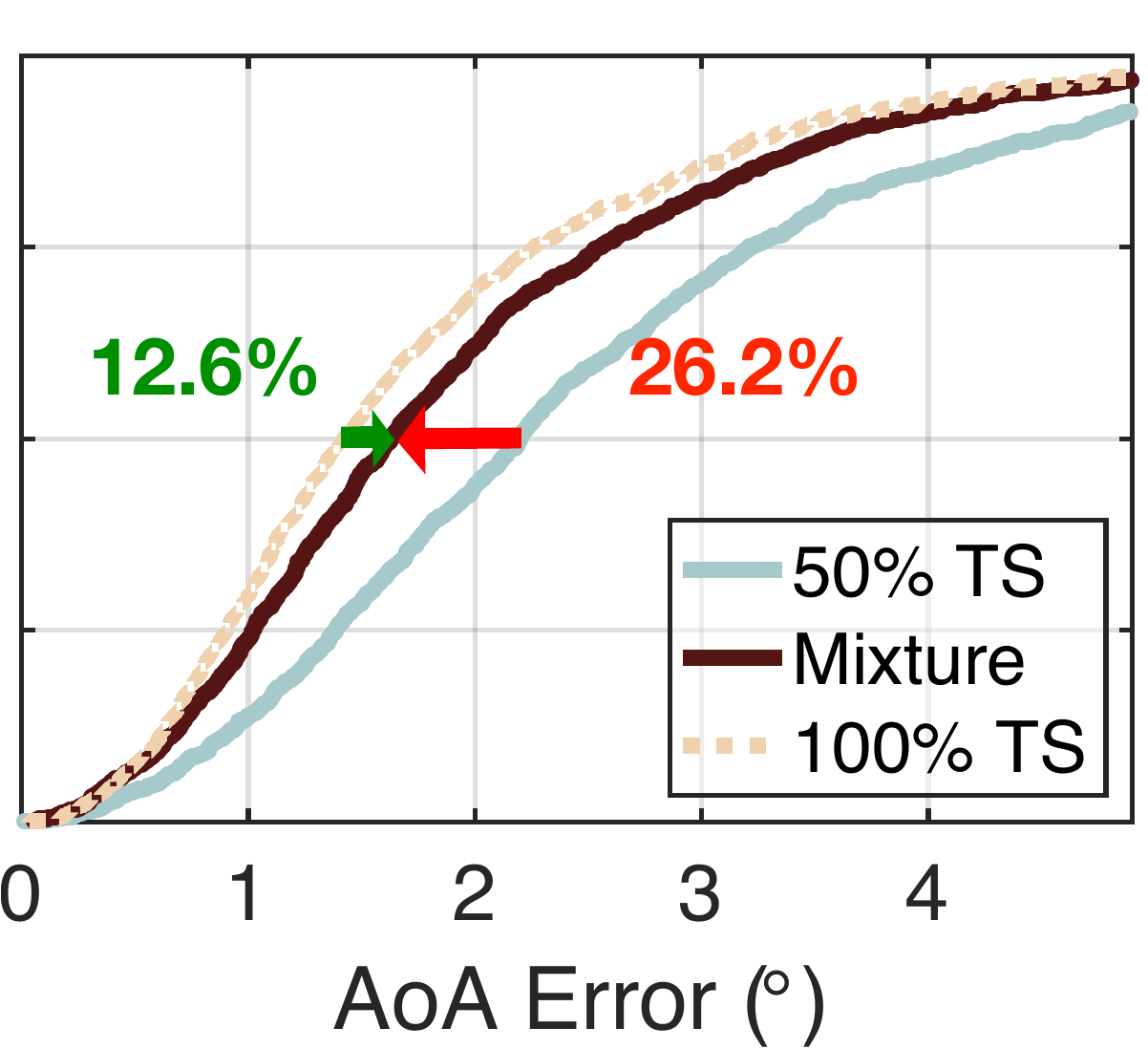}  
   }%
    \subfigure[70\% TS + 30\% SS]{\label{fig:exp-s23-70}
    \includegraphics[height=3cm]{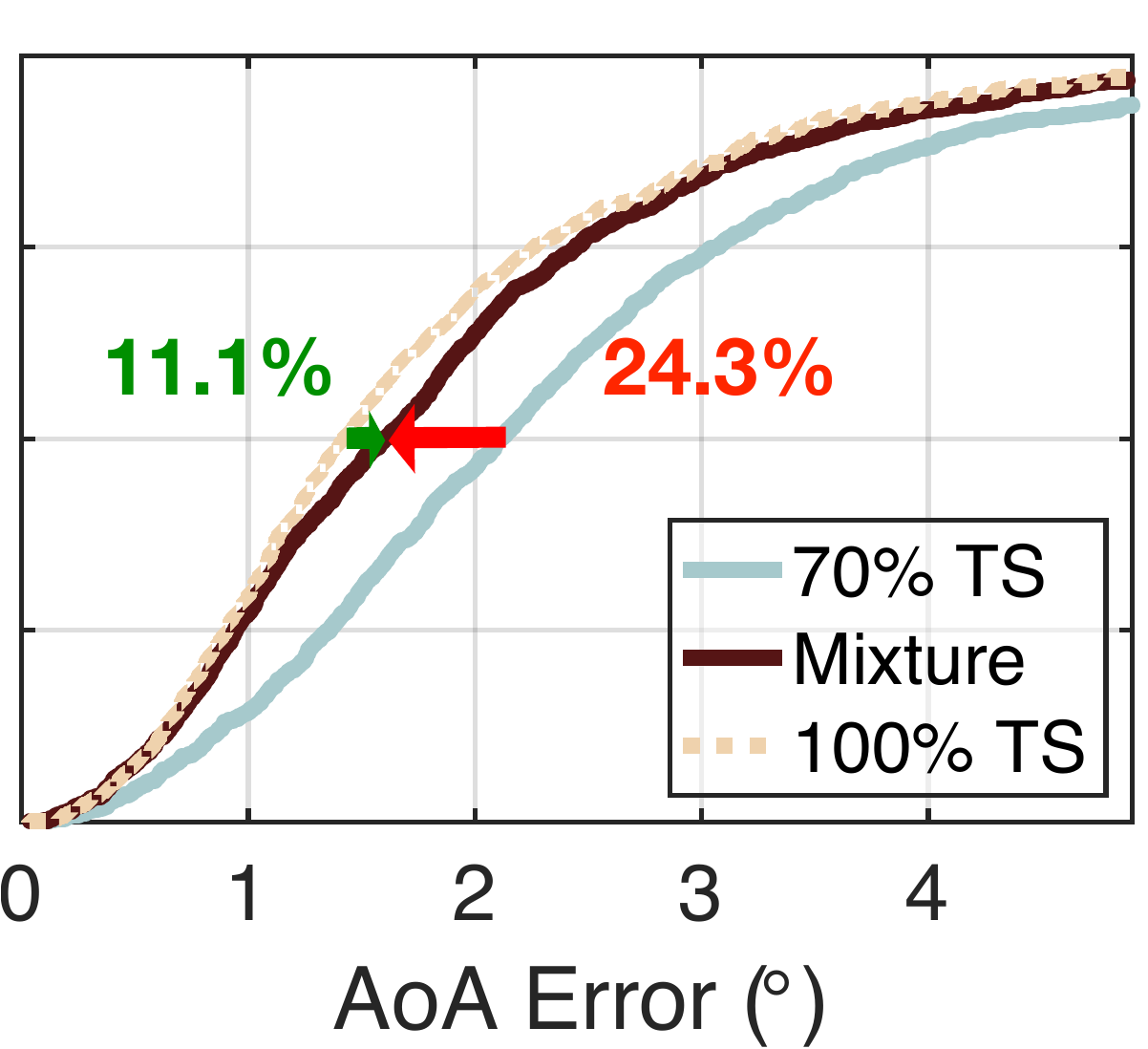}  
   }%
    \subfigure[90\% TS + 10\% SS]{\label{fig:exp-s23-90}
    \includegraphics[height=3cm]{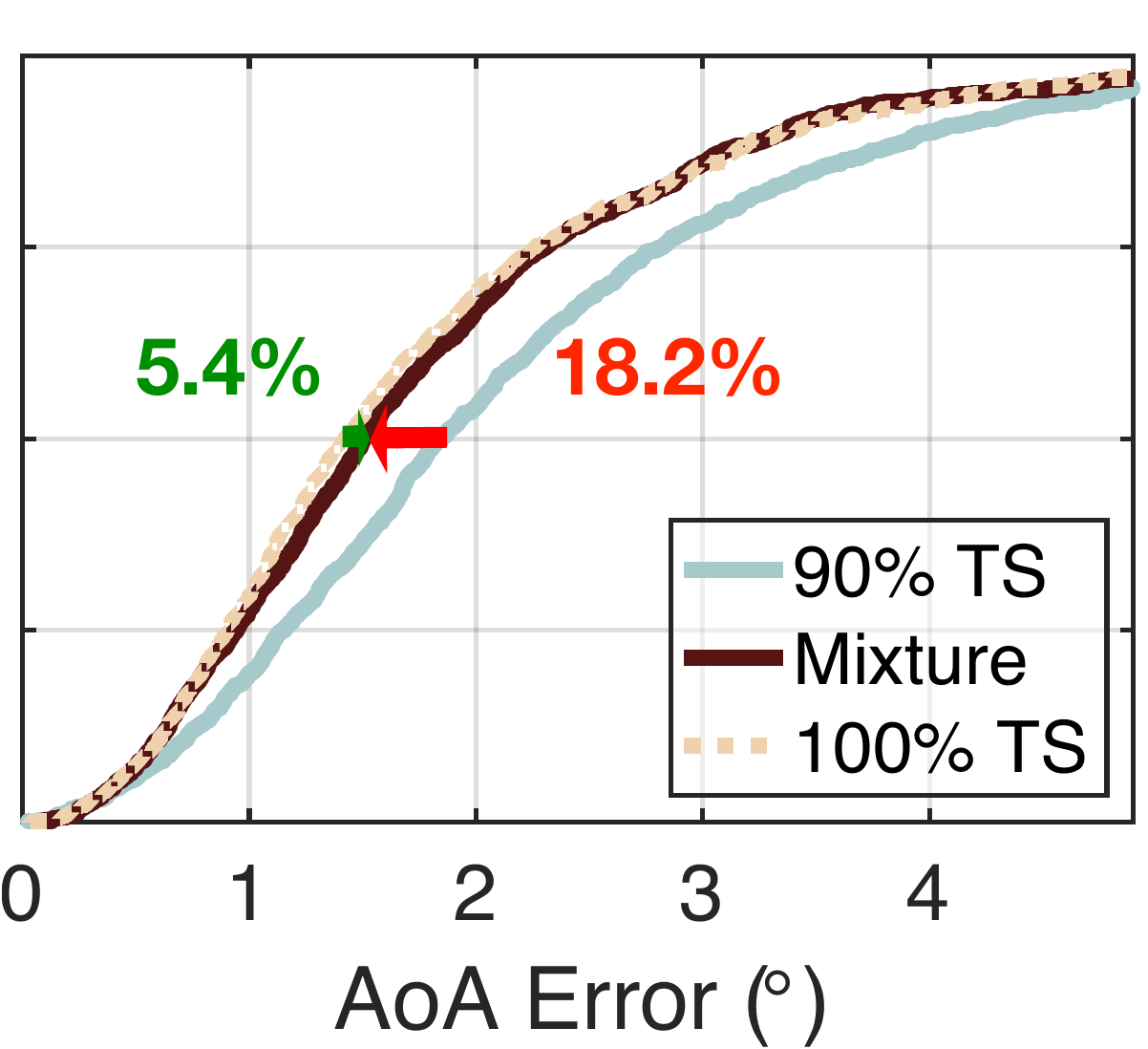}  
   }%
   \vspace{-0.5cm}
   \caption{ CDFs of AoA error. \textnormal{The ANN is trained by the naive-learning (in light blue) and the turbo-learning (in dark red), respectively. We quantify the benefits of \oursystem with different mixture percentages.}}
   \label{fig:aoa-accuracy}
   \vspace{-0.3cm}
\end{figure*}

\section{N\lowercase{e}RF$^\textbf{2}$ Implementation}
\label{section:implementation}

We train a separate \oursystem for each scene. This requires a dataset of RF signals or spatial spectrums captured in the scene, the corresponding locations of TX and RX, and scene bounds (\ie $\Omega$ and $D$). The location-related parameters are acquired by a high-precision infrared positioning system named OptiTrack~\cite{optitrack}. At each iteration, we make the following optimizations:

\textbf{(1) Positional Encoding}: \oursystem accepts two 3D positions and one 2D direction as the inputs. Following the practice from the optical NeRF, which uses the encoded positions, we also raise the dimensions of the inputs  to L using the subsequent encoding function:
\begin{equation}\scriptsize
E(x)=\left(\sin \left(2^{0} \pi x\right), \cos \left(2^{0} \pi x\right), \cdots, \sin \left(2^{L-1} \pi x\right), \cos \left(2^{L-1} \pi x\right)\right)
\end{equation}
This function is applied separately to each of the three coordinate values in the $P_\text{TX}$ or the $P_x$, and to the three components of the Cartesian direction unit vector $\omega$. In our experiments, we set $L=10$ for $P_\text{TX}$ and $P_x$, and $L=4$ for $\omega$. 

\textbf{(2) Voxel Size}: There is a trade-off in setting the size of the voxel. On the one hand, fine-grained voxels can provide higher resolution for \oursystem and accuracy in ray tracing. On the other hand, the number of voxels has a major impact on computational complexity. In our experiments, we set the size of the voxel to the $1/8$ of the wavelength. 

\textbf{(3) Network Configuration}: In each dataset, we randomly sample 80\% samples to train the neural network and use the remained 20\% for testing.  We adopt a similar configuration as NeRF~\cite{mildenhall2020nerf}. Specifically, the batch size is set to 4096. The Adam optimizer~\cite{kingma2014adam} is adopted. The learning rate begins at $3e-4$ and decreases exponentially to $3e-5$. Other hyper-parameters remain at default values (\eg $\beta_1=0.9$, $\beta_2=0.999$, and $\varepsilon=10^{-7}$). The network training for a single scene typically takes around $300$-$500k$ iterations to converge on a single NVIDIA 3080Ti GPU (about 10 hours). In contrast, testing for a single sample can be accomplished in about 0.2 seconds.

\section{Microbenchmark}
\label{section:microbenchmark}

We start with a microbenchmark experiment to provide insights into the working of \oursystem in this section.

\subsection{Experimental Setup}
We deploy a USRP-based RX equipped with a $4\times 4$ antenna array. The RX operates at 915 MHz and targets to receive the signal backscattered from a moving RFID tag. The RFID tag is activated by a nearby reader (\ie 1 m away) and repeatedly transmits \texttt{RN16} replies. Figs.~\ref{fig:predication}-(a) and (b) show the photo of the scene and the corresponding 3D model (composed of a point cloud) scanned by LiDAR. This is a demo room full of reflectors such as metal desks, shelves, tables, computers, and so on. The dataset is created by placing the tag at random positions. For each position, the antenna array generates a spatial spectrum using~Eqn.~\ref{eqn:relative-power}, which is represented by $360\times 90$ pixels from viewpoints sampled on the front hemisphere of the antenna array. We collected a total of 10 K data in this scene, where 8 K are used for training and 2 K for testing. We use the approach introduced in~\cite{miesen2013360} to estimate the phase and amplitude of the received backscatter signals and employ Eqn.~\ref{eqn:case2} as the loss function to train the neural radiance field. 

\begin{figure}[!t]
		\centering
		\includegraphics[width=0.75\linewidth]{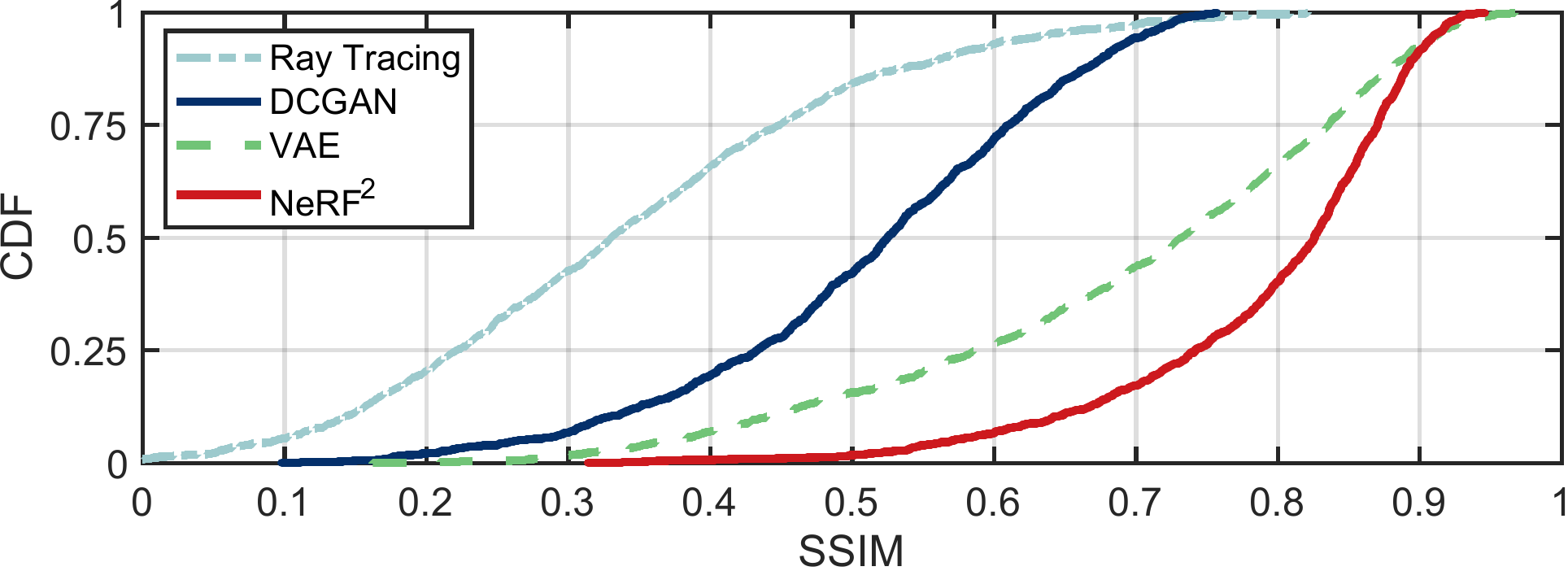}
	    \vspace{-0.3cm} 
	    \caption{SSIM Comparison}
	    \label{fig:exp-rfid-ssim}
	\vspace{-0.8cm}
\end{figure}

\subsection{Spectrum Synthesis}

The goal of the original optical NeRF is to synthesize the photo of the scene taken from an arbitrary direction. Similarly, \oursystem possesses the ability to synthesize RF spatial spectrums when the TX is located at an arbitrary position. To visually understand such a purpose,  we leverage \oursystem to synthesize the spatial spectrums that the antenna array receives. The synthesized spatial spectrum helps us intuitively verify whether the neural radiance field can successfully predict the signal propagations in the scene. We compare \oursystem with the other four baseline schemes. 

\begin{figure*}
	\includegraphics[width=\linewidth]{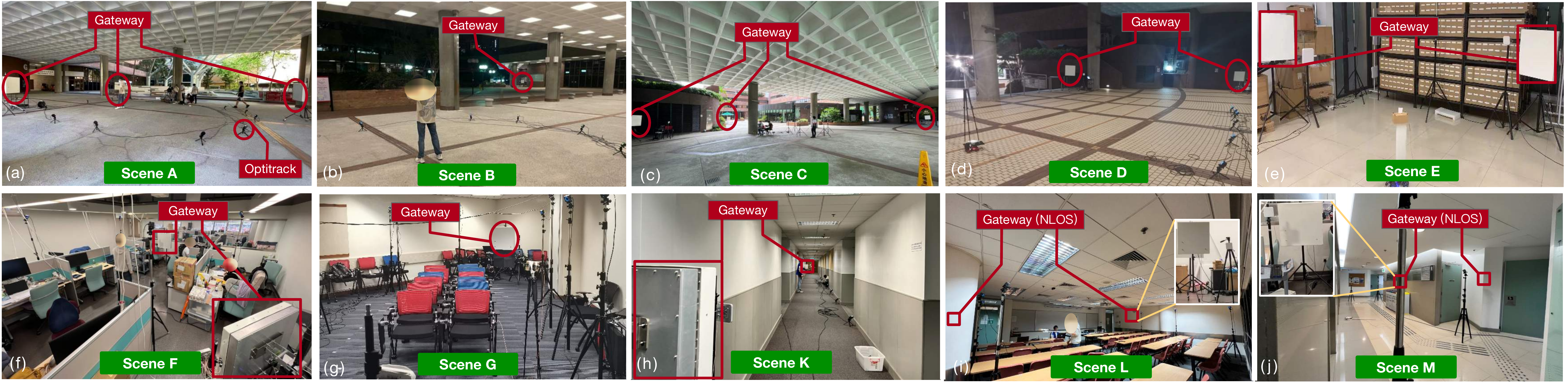}\vspace{-0.3cm}
	\caption{Illustration of example scenes. \textnormal{(a)-(d) shows the semi-indoor environment, which is large-sized and semi-closed halls. (e)-(j) show the full-indoor environment.}}
	\label{fig:scenarios}
	\vspace{-0.5cm}
\end{figure*}

\begin{itemize}[leftmargin=*]	
  \setlength{\parskip}{0pt}
  \setlength{\itemsep}{0pt plus 1pt}
  \item \textbf{Ground truth}: The true spatial spectrums are computed by using the Eqn.~\ref{eqn:spatial-spectrum} across the real signals received by the antenna array. The spectrums are desired to peak at the LOS direction. Unfortunately, Fig.~\ref{fig:predication}-(c) (1st column) shows possible multiple peaks because of the multipath propagations in such a complex environment. 
  \item \textbf{RayTracing}: We employ the RayTracking toolbox in Matlab~\cite{raytracing} to generate the spatial spectrums. Particularly, this toolbox requires importing the 3D model of the scene (\ie Fig.~\ref{fig:predication}-(b)). Given the locations of TX, the toolbox can predict the RF signals received by the RX. 
  \item \textbf{Deep Convolutional Generative Adversarial Network (DCGAN)}. DCGAN is one of the most popular GANs wherein two models (\ie generator and discriminator) are trained simultaneously by an adversarial process. The generator model spawns ``fake'' images that look like the training images. The discriminator model determines whether an image is a real training image or a fake image from the generator. We view the predicted spatial spectrums as images and use DCGAN to learn and generate the spectrums with given TX's locations.    
  \item \textbf{Variational Autoencoder (VAE)}. VAE is one of the famous generative models. It is used to resolve similar issues in wireless systems, such as liquid sensing~\cite{ha2020food} and channel estimation~\cite{liu2021fire}. Adopting the similar architecture in FIRE~\cite{liu2021fire}, an encoder network learns the probability distribution of the training set in a lower dimensional latent space. Subsequently, the samples drawn from the decoder network are decoded to generate the data in accord with the learned distribution.
\end{itemize}

\noindent The results are shown in Fig.~\ref{fig:predication}-(c), where the spatial spectrums are generated using the above schemes when the TX locates at four positions. Visually, the spatial spectrums generated by \oursystem are evidently more similar to the ground truth than other generative models. We further use a common criterion called \emph{structural similarity index measure} (SSIM) to quantify the similarity of two images. Owing to the page limit, we omit the definition of SSIM but encourage the reader to refer to~\cite{wang2004image} for details. A higher SSIM indicates the two images are more similar. We randomly choose 100 positions to synthesize the spatial spectrums using the four algorithms. The CDF of the SSIM between those synthetic spatial spectrums and the ground truth is shown in Fig.~\ref{fig:exp-rfid-ssim}. Particularly, the median SSIM of RayTracking, DCGAN, VAE, and \oursystem are 0.33, 0.52, 0.73, and 0.82, respectively, and their 90th percentiles are 0.56, 0.67, 0.89, and 0.91. The RayTracing underperforms because it is short of the material information, even though the geometric model of the scene is provided. DCGAN and VAE view spatial spectrums as a kind of signature related to the TX's location, so they do not really ``understand'' the rationale behind it. The outperformance of \oursystem is in the accurate model of radiance field in accordance with the underlying physical laws.

\subsection{Performance of Turbo-Learning}
\label{sec:AANN}

\begin{figure}[!t]
		\centering
		\includegraphics[width=\linewidth]{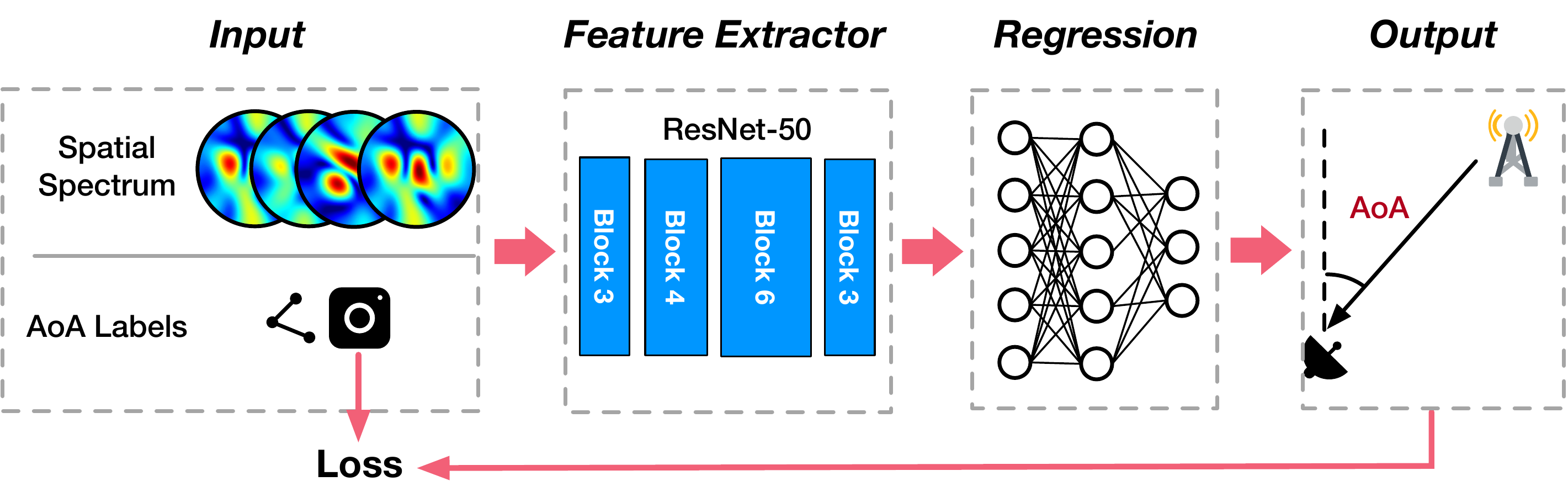}
	    \vspace{-0.7cm} 
	    \caption{Architecture of Angular Artificial Neural Network}
	    \label{fig:ann}
	\vspace{-0.5cm}
\end{figure}

To quantify the benefits of \oursystem, we apply turbo-learning to the AoA estimation, which aims to determine the direction of the line-of-sight propagation. The AoA is desired to be achieved at the peak of the spatial spectrum. Unfortunately, owing to the multipath propagations and the destructive superposition of signals, the peak deviates substantially from the true LOS direction. To address this issue, angular artificial neural networks (AANNs) are resorted to identifying the AoAs~\cite{ayyalasomayajula2020deep,an2020general,babakhani2021bluetooth,comiter2018localization}. Similar to the iArk~\cite{an2020general}, we set up an AANN based on the ResNet convolutional network~\cite{he2016deep}, as shown in Fig.~\ref{fig:ann}. The AANN accepts spatial spectrums in the image format and outputs the AoAs. In the AANN, a ResNet-50 network is adopted as the feature extractor, which is followed by a fully connected network for regression. 
We train the AANN using the following two approaches:
\begin{itemize}[leftmargin=*]	
  \setlength{\parskip}{0pt}
  \setlength{\itemsep}{0pt plus 1pt}
  \item \textbf{Naive Learning}. We use 10\% of the true training dataset (TS, total 8 K) to train the AANN straightforwardly. In this approach, \oursystem is not involved. 
  \item \textbf{Turbo-Learning}. We use the same $10\%$ of the true dataset to train the \oursystem. Then, we use the well-trained \oursystem to generate the rest $90\%$ synthetic dataset (SS). Finally, the  $10\%$ true dataset and the $90\%$ synthetic dataset (\ie turbocharger) are mixed to train the AANN.   
\end{itemize}

\noindent These two learning approaches fully use the same 10\% of the true training dataset for the sake of fairness, \ie \emph{both hold the same amount of information from the true dataset.}  We also use $100\%$ training set to train the AANN as the baseline. 
The results are shown in Fig.~\ref{fig:aoa-accuracy-10}. The median errors of naive learning and turbo-learning are $3.78^\circ$ and $1.96^\circ$, respectively. 
The result of naive learning is enhanced by \oursystem with 47.9\%. On the other hand, the accuracy of turbo-learning is extremely approaching the $1.42^\circ$ error that the baseline achieves. This result demonstrates that the quality of the synthetic dataset generated by \oursystem is as good as the true dataset. Clearly, the quantity of true training dataset required by turbo-learning is far less than the baseline, but the accuracy remains at a comparably high level. This feature is useful because collecting a training dataset is an important but cumbersome and painful task for today's deep learning. The power of \oursystem is in the significant reduction of the quantity of true training set and the corresponding workload.

We also test other mixture ratios (30\% TS+70\% SS, 50\% TS+50\% SS, 70\% TS+30\% SS, and 90\% TS+10\% SS) using the same ways. The results are shown in Fig.~\ref{fig:aoa-accuracy}(b)-(e). As desired, the error of turbo-learning is reduced from $1.96^\circ$ to $1.72^\circ$, $1.62^\circ$, $1.59^\circ$, and $1.50^\circ$. Evidently, the accuracy is increased with an increasing quantity of true datasets. This is understandable because the accuracies of \oursystem and AANN improve as the amount of true information increases. On the other hand, turbo-learning outperforms naive learning by 47.9\% to 29.1\%, 26.2\%, 24.3\%, and 18.2\%. This demonstrates that more benefits can be gained when more percent of synthetic data is given. Even if only 10\% synthetic data is fed, the median error can be reduced by 18.2\% compared with naive learning. One may wonder why do not try the mixture of 0\% TS plus 100\% SS. It is impossible because the training of \oursystem must require a few numbers of true datasets. To achieve the trade-off between the accuracy and the quantity,  it is advisable to take the mixture of 30\% TS plus 70\% SS in practice.

\subsection{Large-scale Experiments}

\begin{table}[]
\linespread{0.9}
\newcommand{\tabincell}[2]{\begin{tabular}{@{}#1@{}}#2\end{tabular}}
\footnotesize
\caption{Summary of Experiment Scenes}
\label{tab:ray-summary}
\vspace{-0.3cm}\scriptsize
\begin{threeparttable}
\begin{tabular}{|c|c|c|c|l|c|c|}
\hline
\rowcolor{black}                                                             \color{white} \tabincell{c}{\textbf{Env.} \\ (\#)}                        &\color{white} \tabincell{c}{\textbf{Scene} \\ (\#)}                             &\color{white} \tabincell{c}{\textbf{RSS} \\(dBm)}                       &\color{white} \tabincell{c}{\textbf{Total}    \\ (\#)}                    &\color{white} \tabincell{c}{\textbf{Density} \\ $(\mathrm{p} / m^{3})$}   &\color{white} \tabincell{c}{\textbf{Space} \\ $(m^{2})$}                &\multicolumn{1}{c|}{\color{white} \tabincell{c}{\textbf{Distance} \\ $(m)$}} \\ \hline

\multirow{4}{*}{\begin{sideways}\textbf{Semi}\end{sideways}}     & \textbf{A}       & $-62.5$      & 84,392       & 3,843.0      & 78.5       & 5         \\  \cline{2-7}                    
 
&                                                                              \textbf{B}       & $-88.6$      & 50,186       & 10,490.4        & 7,854.0	  & 50        \\  \cline{2-7}

&                                										       \textbf{C}       & $-68.3$      & 18,726       & 6,079.9      & 1,256.6    & 20        \\  \cline{2-7}

&                                                                              \textbf{D}       & $-68.9$      & 77,538       & 4,345.1      & 530.9	      & 13        \\  \cline{1-7}

\multirow{10}{*}{\begin{sideways}\textbf{Full}\end{sideways}}    & \textbf{E}       & $-66.2$      & 78,635       & 27,924.4     & 314.2	      & 10        \\  \cline{2-7}

&                                                                              \textbf{F}       & $-65.1$      & 48,467       & 22,627.0     & 153.9	   & 7         \\  \cline{2-7}

&                                                                              \textbf{G}    	 & $-61.4$      & 10,521       & 4,911.8      & 78.5		   & 5         \\  \cline{2-7}

&                                                                              \textbf{H}       & $-61.7$      & 5,102        & 912.7        & 113.1	   & 6         \\  \cline{2-7}

&                                                                              \textbf{I}		 & $-60.9$      & 7,466        & 823.0        & 113.1	   & 6         \\  \cline{2-7}

&                                                                              \textbf{J}   	 & $-61.6$      & 25,543       & 1,576.7      & 78.5		   & 5         \\  \cline{2-7}

&                                                                              \textbf{K}   	 & $-71.0$      & 28,882       & 1,380.6      & 1256.6	   & 20        \\  \cline{2-7}

&                                   											\textbf{L}   	 & $-77.7$      & 52,634       & 935.9        & 1963.5	   & 25        \\  \cline{2-7}

&                                   											\textbf{M}       & $-79.9$      & 21,683       & 1,335.2      & 3217.0	   & 32        \\  \cline{2-7}

&                                   											\textbf{N}       & $-68.5$      & 21,729       & 848.1        & 254.5	   & 9         \\
\bottomrule[1pt]                         
\end{tabular}
\end{threeparttable}
\vspace{-0.6cm}
\end{table}

Regardless of NeRF or \oursystem, both are scene-dependent because the radiance field is highly related to the scene layout. Whether the outperformance of turbo-learning can still be achieved in different scenes is unclear. Thus, we conduct large-scale experiments. We use the same antenna array to collect a huge dataset at 531,504 positions from 14 scenes (labeled A$\sim$N). The settings are listed in Table~\ref{tab:ray-summary}. Fig.~\ref{fig:scenarios} shows eight of them (owing to the space limit). We first collect the data in a large-area semi-indoor environment with the purpose of quantifying the impact of the distance. In such an environment, the antenna array is deployed in four scenes labeled A, B, C, and D, which are large-area and semi-closed halls, as shown in (a)-(d) of Fig.~\ref{fig:scenarios}. In these scenes, the distance varies from 5 to 50 m. The distance is the mean value between the scene center and the antenna array. We then collect the data in the full-indoor environment. We deploy the platform in 10 rooms (\ie Scenes E-N). Scene E is a warehouse, Scene F is a lab room, and Scenes G-I are classrooms. Scenes J and N are offices, Scene K is the hallway, Scene L is a meeting room, and Scene M is a lift lobby, as shown in (e)-(j) of Fig.~\ref{fig:scenarios}. The coverage of the scene ranges from  5 to 32 m. Particularly, the gateways are deployed behind the wall in Scenes L, M, and N. Majority of these data are collected in the scenes full of people passing by and various reflectors.

\begin{figure}[!t]
		\centering
		\includegraphics[width=0.85\linewidth]{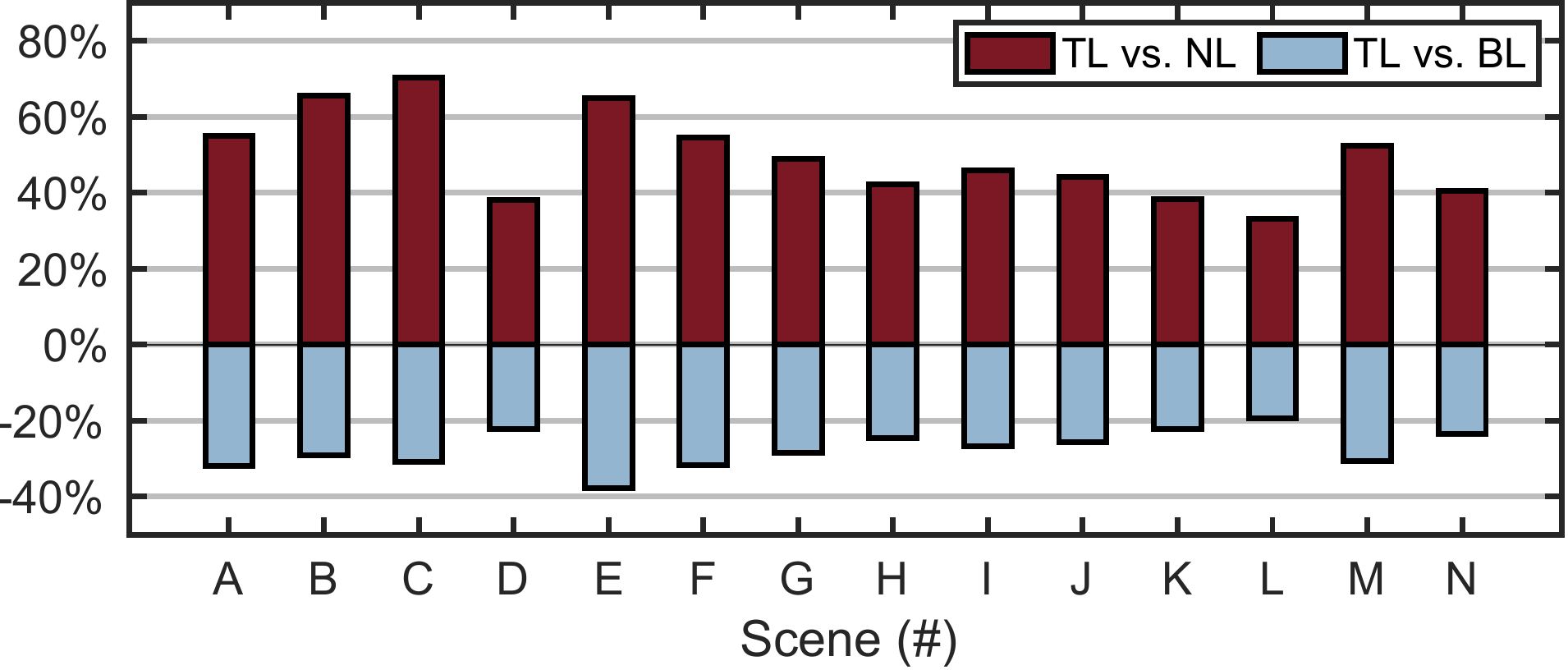}
	    \vspace{-0.4cm} 
	    \caption{AoA Accuracy vs. Scenes}
	    \label{fig:aoa-scene}
	\vspace{-0.6cm}
\end{figure}

Similarly, we choose the 80\% dataset for training and the 20\% dataset for testing in each scene. Naive learning with the entire 80\% true dataset is used for the baseline. Turbo-learning is conducted with 10\% (out of the 80\%) of the true training dataset plus 90\% synthetic dataset. The AoA accuracy results are shown in Fig.~\ref{fig:aoa-scene}. From the figure, we have the following two findings:
\begin{itemize}[leftmargin=*]	
  \setlength{\parskip}{0pt}
  \setlength{\itemsep}{0pt plus 1pt}
  \item Compared with naive learning (NL), the turbo-learning (TL) can offer  $33\%$-$70\%$ improvement. The average is 49.5\%. This shows that the performance enhancement by turbo-learning is a general phenomenon across scenes. 
  \item Compared with the baseline (BL),  the turbo-learning can hold $-27.5$\% gap, where the minus sign denotes the ``lower accuracy than''. However, turbo-learning saves 90\% workload for the dataset collection because only 10\% training set is used. 
\end{itemize}  

Our experiments reveal two key factors influencing turbo-learning performance: (1) Quantity of the dataset. \oursystem has the ability to reduce the requirement for data collection in application-layer NN tasks. However, if adequate data is provided, the application-layer NNs can train the model effectively, thus reducing the benefits of \oursystem. (2) Quality of the dataset. The performance of \oursystem can also be affected by environmental interference, such as passing by people or other signals. Despite this, our results demonstrate that turbo-learning still improves the performance of application-layer NNs by over 30\%. In summary,  the outperformance of turbo-learning is mainly derived from the physical model provided by \oursystem. Naive learning models the ``signature (feature)-based'' relationship between the AoA and the spectrums, but \oursystem learns the physical rationale behind the relationship so it can provide more reasonable samples for the learning.

\section{Field Study: BLE Localization}
\label{section:localization}

In this section, we discuss how \oursystem helps indoor localization in the scenario where no antenna array is available at a receiver. We conduct a large-scale experiment with 50 BLE gateways in an elderly nursing home. The project aims to track the potential spread of COVID-19 to protect elderlies from the infection better.    

\subsection{Experiment Setup}

\begin{figure}[!t]
	\centering
	\includegraphics[width=0.85\linewidth]{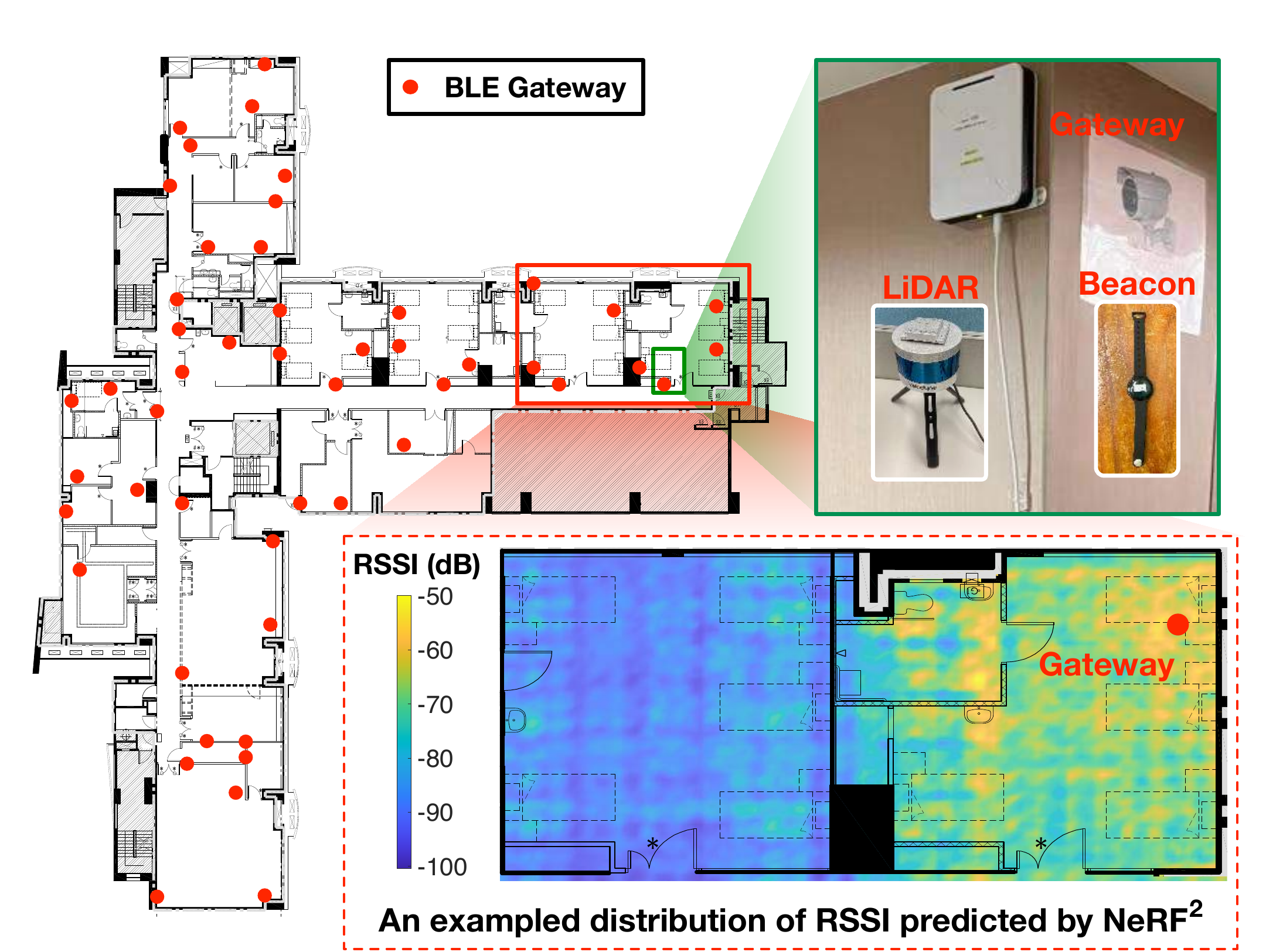}
	\vspace{-0.2cm} 
	\caption{The floor plan of the nursing home and deployment of BLE gateways.}
	\label{fig:floor-plan}
	\vspace{-0.4cm}
\end{figure}

Fig.~\ref{fig:floor-plan} shows the floor plan of the facility, which occupies 15,000 ft$^2$. A total of  50 BLE gateways (red circles) are deployed to collect the ID and RSSI of  BLE beacons. Each gateway is $72 \times 7\times 20 \text{mm}^3 $ in size, operates at 2.4 GHz and adopts an NRF52832 Bluetooth SoC~\cite{ble-soc} from Nordic Semiconductor. Redundant gateways are deployed to ensure that each location can be covered by at least 3 gateways. The BLE nodes are embedded into the visitor cards or elderlies' wristbands. They broadcast every 500 ms with 4 dBm transmitting power.

\textbf{Ground Truth.}  The Velodyne VLP-16 LiDAR plus a 9-axis IMU are used to serve LIO-SAM (\ie a publicly available SLAM algorithm~\cite{liosam2020shan}) for localization and map construction. The gateways and nodes are located by the LiDAR system as the ground truth. Taking 30 BLE nodes, we randomly walk into the house and totally create a dataset involving 6~K positions in the scene. Each dataset item is a 50-dimensional tuple, including the RSSI values detected by the 50 gateways, plus the position of the BLE node. The RSSI value is set to -100 dB by default if the gateway does not detect any signal from the node. 70\% (4.2 K) and 30\% (1.8 K) of the dataset are chosen from training and testing datasets.

\subsection{RSSI Prediction}

\begin{figure}
  \begin{minipage}[t]{0.5\linewidth}
 	\centering
	\includegraphics[width=1\linewidth]{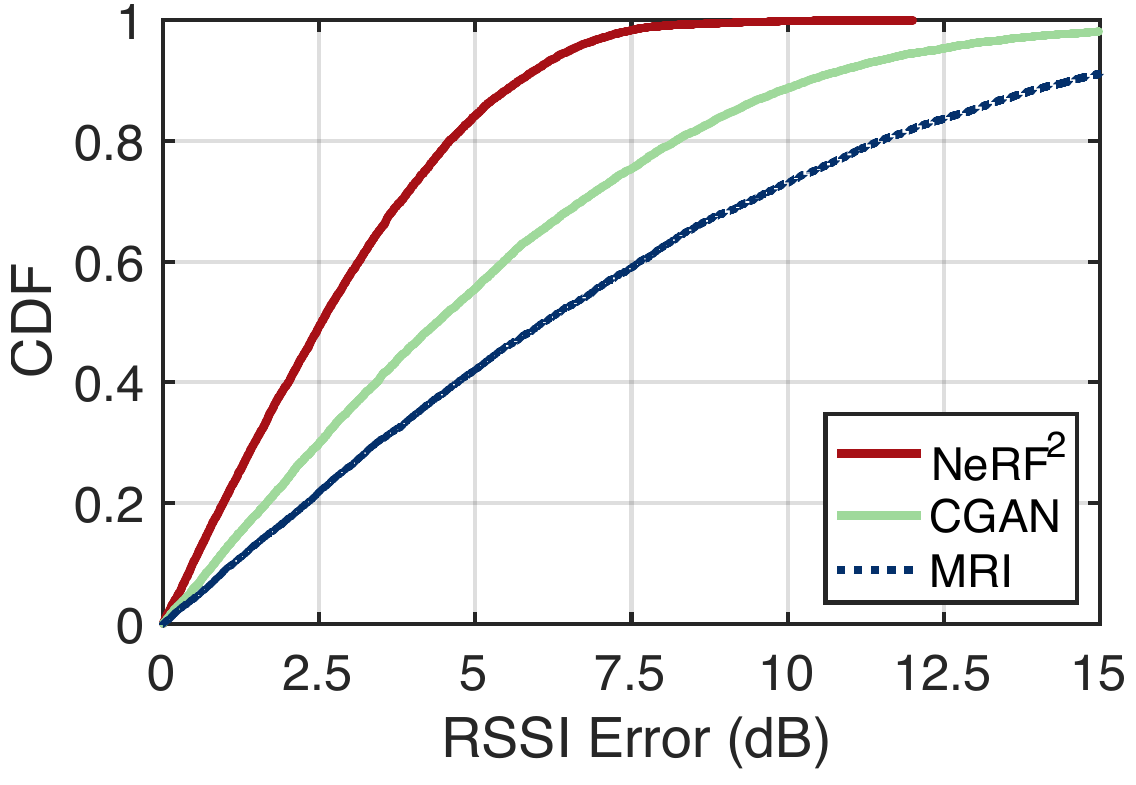}
	\vspace{-0.8cm} 
	\caption{RSSI Prediction}
	\label{fig:exp-ble-rssi}
  \end{minipage}%
  \begin{minipage}[t]{0.5\linewidth}
 	\centering
	\includegraphics[width=1\linewidth]{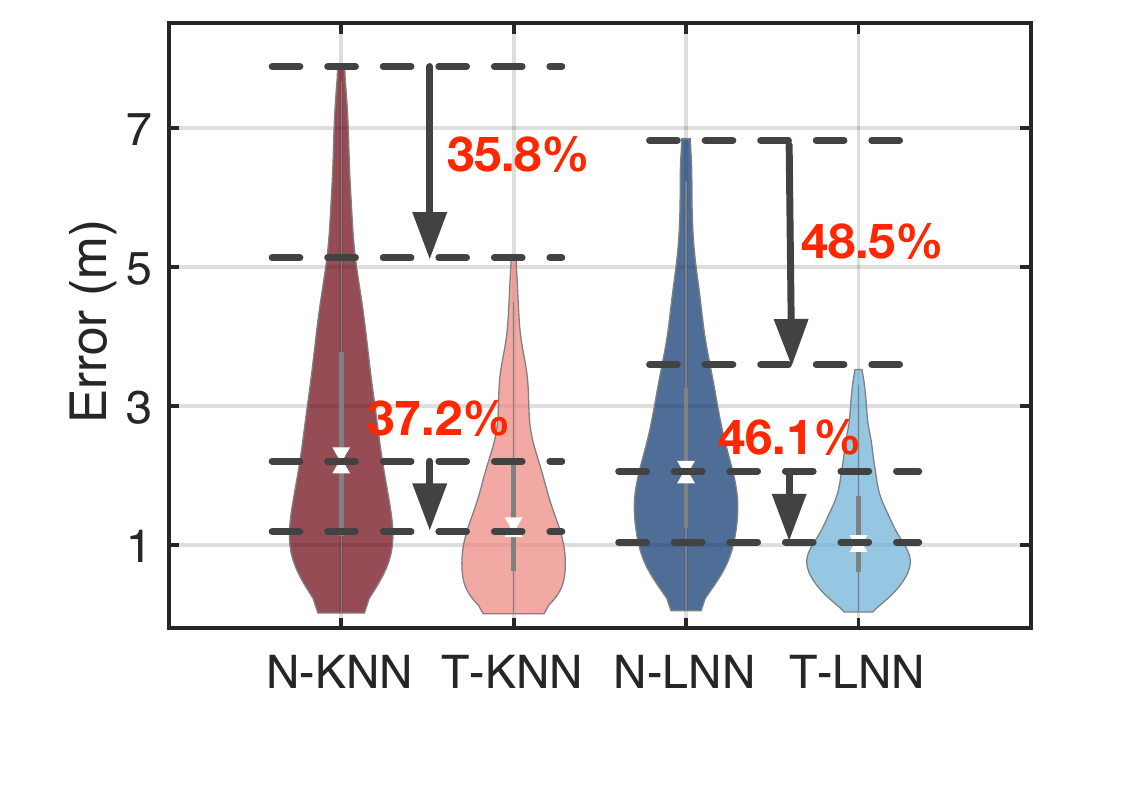}
	\vspace{-0.8cm} 
	\caption{Localization Result}
	\label{fig:exp-ble-loc}
  \end{minipage}
  
    \begin{minipage}[t]{1\linewidth}
 	\centering
	\includegraphics[width=1\linewidth]{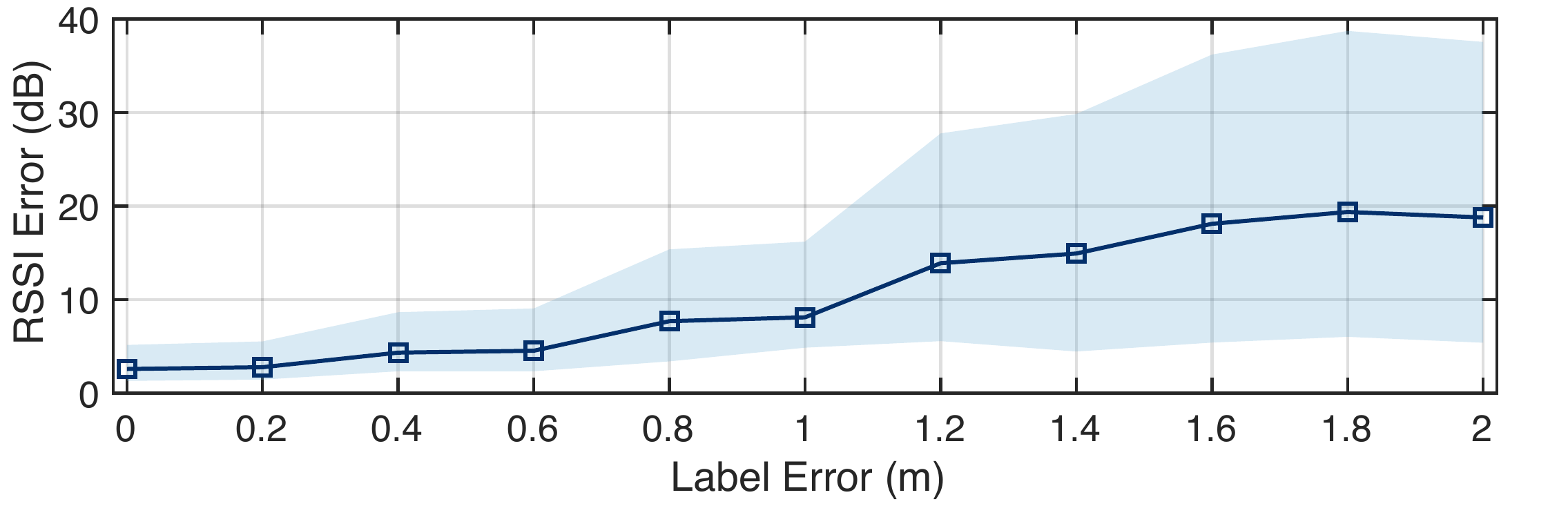}
	\vspace{-0.8cm} 
	\caption{RSSI vs. Label error}
	\label{fig:exp-ble-label}
	\vspace{-0.5cm}
  \end{minipage}\vspace{-0.4cm}
\end{figure}

Data-driven approaches are emerging as promising solutions for BLE localization, such as KNN, SVM, and MLP. These methods require an accurate dataset for fingerprint matching or network training. Here, we apply turbo-learning for BLE localization, where the \oursystem is trained using the single-antenna RX model (see $\S$\ref{section:single-antenna}). The training process is more complicated than in previous cases in that we have 50 RXs here. The beacon of the same BLE node may be received by multiple gateways simultaneously. In this case, we must take ray tracing multiple times,  in each of which the result is an aggregation of signals arriving at the corresponding RX from all possible directions (Eqn.~\ref{eqn:single-antenan-model}). Similarly,  given a position that a BLE node locates in the scene, we must take the ray tracing to predict the RSSI of the signal received at any gateway with the help of \oursystem. Fig.~\ref{fig:floor-plan} shows an example distribution of the predicted RSSI across the two rightmost rooms. It can be seen that the coverage of a gateway is not as good as that the manual claims (\ie 10 m). The signal becomes very weak after walls. Thus, we deployed 3-4 gateways in each room to ensure full coverage. For comparison,  we also adopt two other proposed prediction approaches, MRI~\cite{shin2014mri} and CGAN~\cite{parralejo2021comparative}. MRI interpolates the RSSI values at the unsampled location using a basic radio propagation model. CGAN uses the conditional generative adversarial network to predict the RSSI values straightforwardly without regarding any physical model. The prediction error is defined as the difference between the predicted and the collected RSSI values at 1.8 K tested positions. The CDFs of the prediction error are shown in Fig.~\ref{fig:exp-ble-rssi}. As a result, the median of \oursystem is 2.6 dB ($10^\text{th}$ percentile: 0.5 dB; $90^\text{th}$ percentile: 5.7 dB). By contrast, the median errors of CGAN and MRI are 4.5 dB and 6.2 dB, respectively. Evidently,  \oursystem performs far better than the two others because it combines the advantages of deep learning (\eg CGAN) and the physical model (\eg MRI). The physical model provides prior knowledge about signal propagation, while deep learning uses statistical models to depict complicated RF interactions.   

\begin{figure*}
	\begin{minipage}[t]{0.24\linewidth}
		\centering
		\includegraphics[width=1\linewidth]{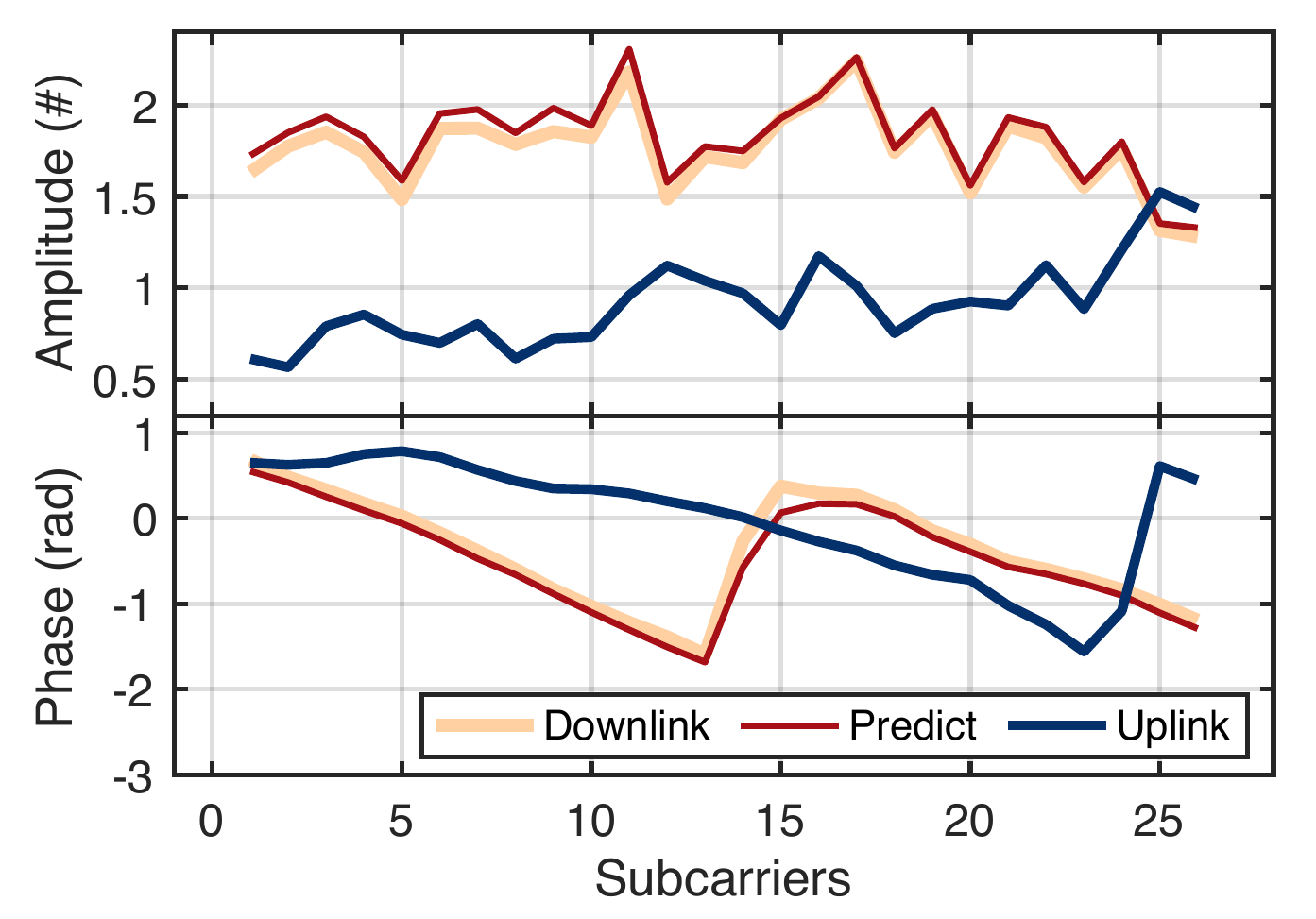}
		\vspace{-0.8cm} 
		\caption{Channel Amplitude \& Phase}\vspace{-0.8cm}
		\label{fig:exp-mimo-ampphs}
	\end{minipage}%
	\begin{minipage}[t]{0.24\linewidth}
		\centering
		\includegraphics[width=1\linewidth]{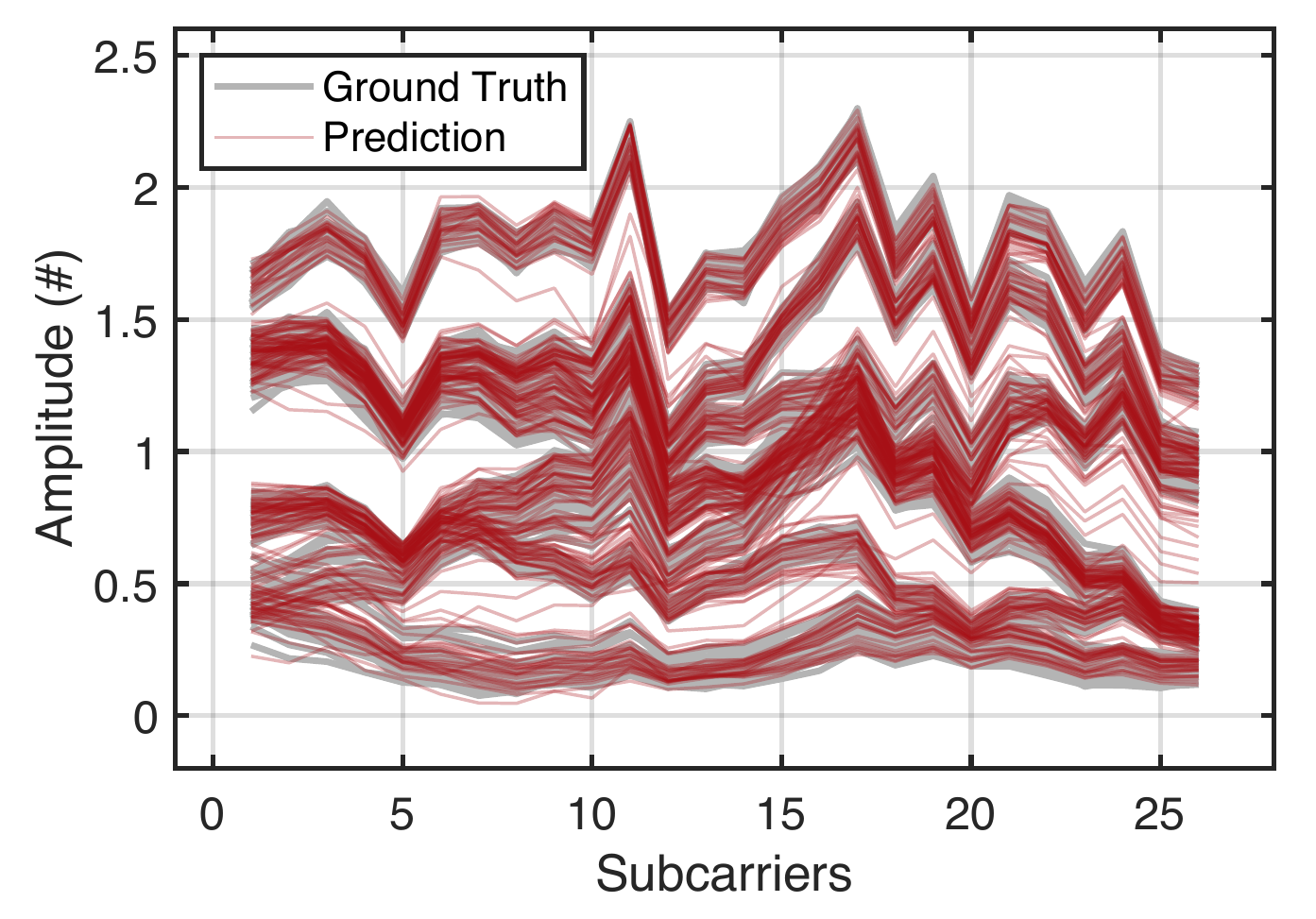}
		\vspace{-0.8cm} 
		\caption{Channel Amplitude in 2s}\vspace{-0.8cm}
		\label{fig:exp-mimo-bfgain}
	\end{minipage}
	\begin{minipage}[t]{0.24\linewidth}
		\centering
		\includegraphics[width=1\linewidth]{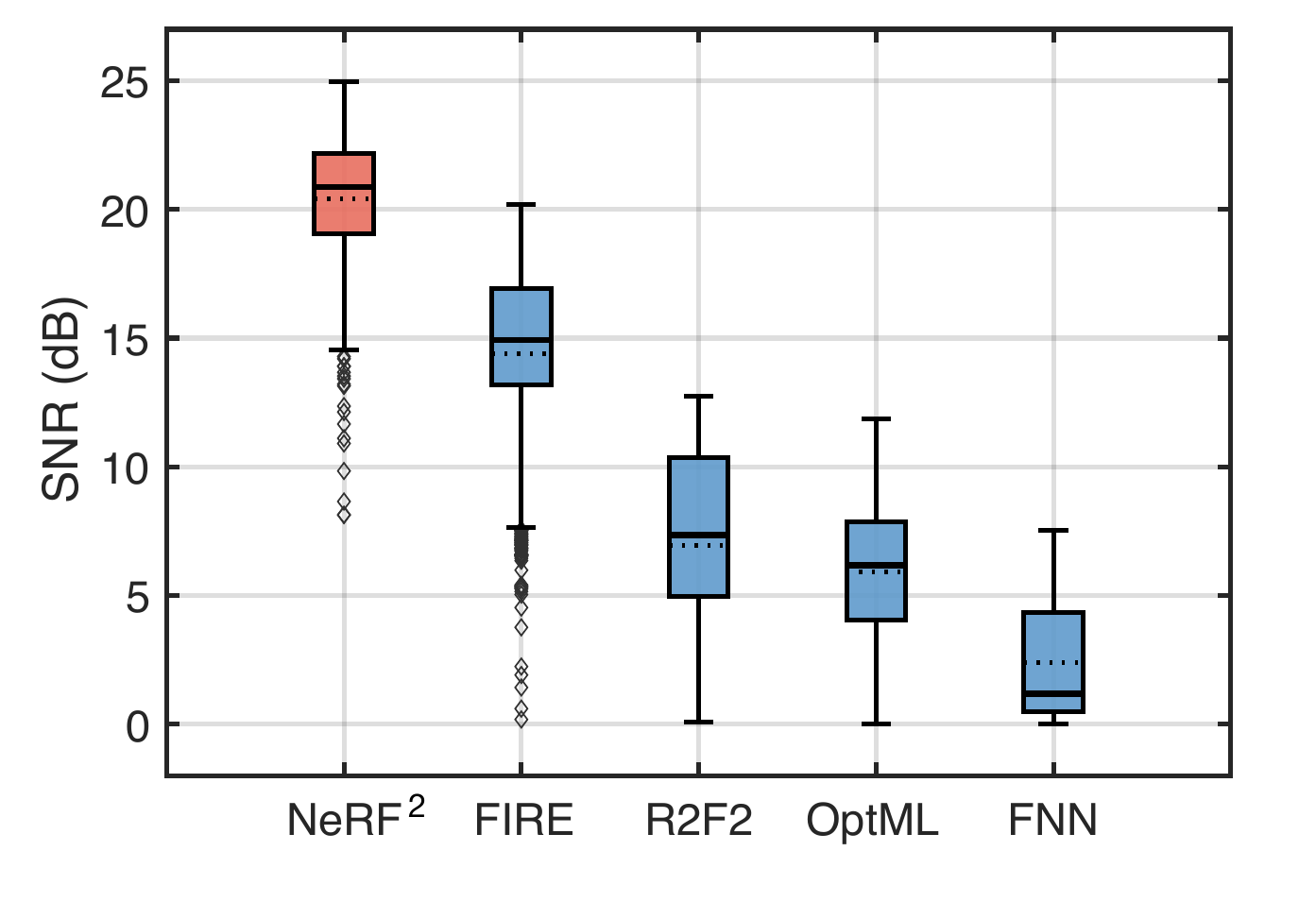}
		\vspace{-0.8cm} 
		\caption{Prediction SNR}\vspace{-0.8cm}
		\label{fig:exp-mimo-compare}
	\end{minipage}
	\begin{minipage}[t]{0.24\linewidth}
		\centering
		\includegraphics[width=1\linewidth,height=3cm]{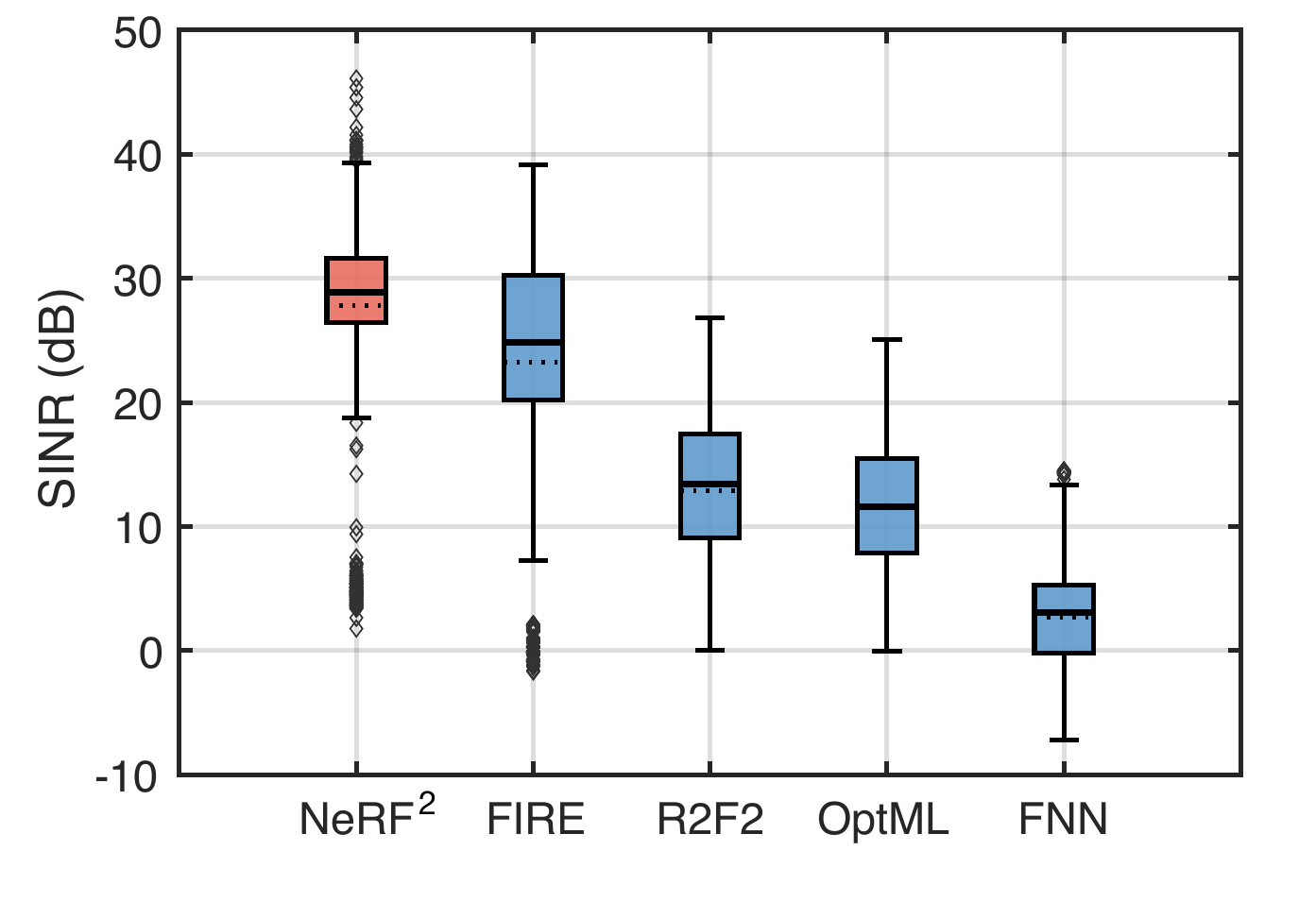}
		\vspace{-0.8cm} 
		\caption{MU-MIMO SINR}\vspace{-0.8cm}
		\label{fig:exp-mimo-sinr}
	\end{minipage}%
\end{figure*}

\subsection{Localization Results}
We use the well-trained \oursystem to generate a 20 K synthetic dataset at random locations and feed them to the following two localization algorithms.

$\blacksquare$ \textbf{Turbo-KNN (T-KNN)}: We first evaluate the fingerprint-based localization approach, which assumes the RSSI values are highly related to a node's location. The $K$-nearest positions are chosen to compute the target node's location, where the RSSI values collected from these $K$ positions (saved in a database) are most close to the RSSI value collected from the unknown position. The node is located at the weighted average of the $K$ positions~\cite{ni2003landmarc}. Fig.~\ref{fig:exp-ble-loc} shows the localization accuracy of naive-KNN (N-KNN) and T-KNN. The N-KNN only adopts the 4.2K true dataset only, whereas T-KNN uses the 20 K synthetic dataset. The median error of  T-KNN  is 1.41 m ($10^\text{th}$ percentile: 0.27 m; $90^\text{th}$ percentile: 3.3 m), whereas that of N-KNN is 2.52 m ($10^\text{th}$ percentile: 0.61 m; $90^\text{th}$ percentile: 5.38 m). Turbo-learning helps the KNN-based localization approach reduce the error by 44\%.

$\blacksquare$ \textbf{Turbo-LNN (T-LNN)}: We build another neural network to learn the mapping between an RSSI tuple and a position. We call this network \emph{localization neural network} (LNN), which accepts the 50-dimensional RSSI tuple as input and outputs the position. The LNN consists of five-layer fully connected layers with the ReLU activation function. Similarly, the LNN is trained by using the 4.2 K true dataset and 20 K synthetic dataset, respectively. We call them naive-LNN (N-LNN) and T-LNN. Fig.~\ref{fig:exp-ble-loc} shows their results. The median error of T-LNN is 1.11~m ($10^\text{th}$ percentile: 0.34 m, $90^\text{th}$ percentile: 3.46 m), whereas that of N-LNN is 2.26 m ($10^\text{th}$ percentile: 0.75 m, $90^\text{th}$ percentile: 6.78 m). The error is reduced by turbo-learning by 50.8\% (\ie 1.2 m error). Particularly, T-LNN further reduces the 30 cm median error than T-KNN.

In summary,  \oursystem-powered turbo-learning can effectively reduce the localization errors by $\sim 50$\% regardless of which data-driven approach is used. It also decreases the standard variance by $\sim 40\%$ because the scale of training data is enlarged by $5\times$ and a larger number of samples clearly benefit the convergence. 

\subsection{Impact of Label Errors}

Label accuracy critically affects deep learning algorithm performance, so we investigate the impact of beacon location label errors on RSSI prediction accuracy. Fig.~\ref{fig:exp-ble-label} shows the results, representing median prediction error and quartiles. We introduce uniformly distributed noise to the location label to simulate errors, whereby the label is reported with the circular error of radius $r$. The error $r$ is increased from 0 m to 2 m in a step of 0.2 m, and the corresponding median RSSI prediction error degrades from 2.6 dB to 18.8 dB, with a near $7\times$ increase. The standard deviation also increases from 1.9 dB to 9 dB. When the label error exceeds 1 m, the prediction accuracy of RSSI decreases severely. This is primarily because when the error surpasses 1 m, the label may be inaccurately situated in another room. On the other hand, when is label error is less than 0.6 m, the RSSI prediction error is below 4.5 dB, which is only 1.9 dB when compared to the absence of error. As such, we adopt the SLAM algorithm, which guarantees a label error of less than 0.2 m, as our preferred method of collecting ground truth data.

\section{Field Study: 5G MIMO}
\label{section:beamforming}

In this section, we discuss how \oursystem helps massive MIMO channel estimation of the Frequency Domain Duplex (FDD) system for 5G, where the uplink and downlink transmissions operate at different frequencies. Therefore, the principle of reciprocity that two link channels are equal no longer holds~\cite{liu2021fire}. To estimate the downlink CSI for beamforming, the client devices must receive extra symbols transmitted by the base station equipped with a massive antenna array and then send the estimated result back to the base station, leading to unsustainable overheads. To solve this problem, substantial research is devoted to predicting the downlink channel state by observing the uplink channel, based on the path-sharing assumption that both link channels are created by the same underlying physical environment and the identical paths being traveled~\cite{vasisht2016eliminating,bakshi2019fast,liu2021fire}. For example, FNN~\cite{bakshi2019fast} and FIRE~\cite{liu2021fire} make use of a fully connected network and a VAE to transfer the estimated CSI from the uplink to the downlink, respectively. 

This problem naturally falls into the domain of \oursystem because \oursystem represents the scene by using an RF radiance field. Given the position of the client device, \oursystem can exactly predict what kind of signal will be received at (or transmitted from) the base station. The key question is how we know the position of the client device. As studied previously~~\cite{kotaru2015spotfi,xie2019md}, the CSI is highly related to the physical environment, so a certain and unique mapping supposedly exists between the position of the client device and the CSI of its uplink signal, which is similar to fingerprint-based localization. Unlike previous works, which attempt to transfer the uplink CSI to the downlink CSI, we use the uplink CSI as a position indicator to train \oursystem, which then predicts the downlink CSI directly. Formally, Eqn.~\ref{eqn:F} is rewritten as:
\begin{equation}\label{eqn:F}\footnotesize
	\mathbf{F}_{\Theta}:  (I_\text{uplink},  \omega, P_x) \rightarrow (\delta_x, S_x) 
\end{equation}
where $I_\text{uplink}$ is the position indicator (\ie uplink CSI). However, the following ray tracing works at the downlink frequencies, and the network is trained with the collected downlink CSI. Thus, our solution does not rely on the path-sharing assumption anymore. 

\vspace{-0.1cm}
\subsection{Experiment Setup}
We choose the publicly available Argos channel dataset~\cite{shepard2016understanding} for our evaluation. Argos dataset is a real-world multi-user MIMO (MU-MIMO) dataset that contains 104 antennas at the base station and eight users, including mobile and static traces collection. The dataset contains two different working frequency versions, \ie 2.4 GHz and 5 GHz. We need to train two separate \oursystem networks for them. The dataset is collected on ArgosV2 platform~\cite{shepard2012argos}, which uses omnidirectional monopole antennas with the spacing of a half wavelength at 2.4~GHz (\ie 63.5~mm). The system has up to 20~MHz bandwidth with 64 OFDM subcarriers. To estimate the CSI, the system sends 802.11 Long Training Symbols pilots at the beginning of each frame with 52 subcarriers. Similar to the previous work~\cite{liu2021fire}, we use the first half of the 52 subcarriers (\ie 26 subcarriers) for the uplink channel, whereas the remained half is the downlink channel. Furthermore, the uplink and downlink channels are separated by guard bands. We aim to predict the downlink CSI from the uplink CSI without any feedback. The dataset is collected in a complex environment with numerous non-line-of-sight propagations. The dataset contains a total of 100 K items; 70\%  are used for the training, and  30\% are used for the testing. We choose  FIRE~\cite{liu2021fire}, R2F2~\cite{vasisht2016eliminating}, OptML~\cite{bakshi2019fast}, and FNN~\cite{huang2019deep} for comparison. For fairness, the experimental results of these four algorithms (including SNR and SINR) are taken from~\cite{liu2021fire}, which are also measured based on the same dataset.  

\vspace{-0.2cm}
\subsection{Channel Prediction Accuracy}

First, we evaluate the accuracy of downlink channel prediction using \oursystem. We show a prediction example in  Fig.~\ref{fig:exp-mimo-ampphs}. The blue uplink CSI, including the phase and amplitude at 26 subcarriers, is fed into \oursystem, and the outputted downlink CSI is shown in yellow. The red ground truth is collected by the real antenna array. The two curves are almost overlapped, exhibiting an incredibly high-accuracy prediction. Fig.~\ref{fig:exp-mimo-bfgain} plots the amplitude of all ground truth and predicted downlink CSI for a mobile client in 2 seconds. Each red curve shows one prediction result of 26 subcarriers at one time. \oursystem can still achieve high-accuracy prediction when the client is moving.
To quantify the accuracy, we adopt the prediction SNR (proposed in~\cite{liu2021fire}), which compares the predicted channel $\textbf{H}$ and the ground truth channel $\textbf{H}_\text{gt}$ as follows:
\begin{equation}\scriptsize
\text{SNR} = -10 \log_{10} \left\|\mathbf{H}-\mathbf{H_{\text{gt}}}\right\|^{2} +10 \log_{10} \left\|\mathbf{H_{\text{gt}}} \right\|^{2}
\end{equation}
A higher positive SNR is obtained when the predicted channel gets closer to the ground truth. Fig.~\ref{fig:exp-mimo-compare} shows the SNR CDF of five prediction algorithms. Clearly,  \oursystem achieves a far higher SNR compared with others. Specifically, \oursystem achieves a median SNR of 20.87 dB ($10^{th}$ percentile: 17.19 dB, $90^{th}$ percentile: 23.14 dB).  The median SNRs of FIRE, R2F2, OptML, and FNN are 14.9 dB, 7.3 dB, 16.1 dB, and 1.2 dB, respectively. \oursystem is  5.97 dB better than the VAE-based FIRE, 13.57 dB better than R2F2, 4.77 dB better than OptML, and 19.67 dB better than FNN. This demonstrates the high-accuracy prediction ability of \oursystem again.

 \subsection{MU-MIMO Performance}
 
Finally, we evaluate the performance of \oursystem for MU-MIMO. In an MU-MIMO system, the base station transmits to multiple clients simultaneously through the separated beamforming. MU-MIMO requires a highly accurate channel estimation because a small error resulting from the CSI estimation for one client device may cause leaked interference to other client devices. Owing to the space limit, we encourage the reader to refer to \cite{liu2021fire,spencer2004zero} for details about how the base station encodes the data to achieve the MU-MIMO. Here, we randomly select two clients to communicate with the base station equipped with 8 antennas, which constructs an $8\times 2$ MU-MIMO system. The signal-to-interference-and-noise-ratio (SINR) is used to indicate the performance of an MU-MIMO system. Fig.~\ref{fig:exp-mimo-sinr} shows the results. \oursystem can achieve a median SINR of 29.22 dB. By contrast, the median error of FIRE, R2F2, and OptML are 24.90 dB, 13.33 dB, and 11.53 dB, respectively. The outperformance of \oursystem in the MU-MIMO is derived from the precise channel estimation.

\section{Related Work} 
\label{section:related-work}

Our work falls under broad channel measurements and wireless localization studies.

\textbf{Optical Neural Radiance Field}. NeRF attracts considerable attention in the field of computer vision with the development of related research. It shows great potential for 3D model and environment reconstruction~\cite{liu2020neural, tancik2022block, rematas2022urban}, scene relighting and view synthesis~\cite{srinivasan2021nerv, martin2021nerf}, and so on. To the best of our knowledge, \oursystem is the first to model the neural radiance fields on the basis of RF signals. Unlike optical NeRF, \oursystem considers the phase of RF signals, which requires a different physical tracing model. Moreover, owing to the limited size of the antenna array compared with the camera, \oursystem proposes two different training approaches. By precisely predicting the RF signal, \oursystem benefits many RF applications, such as wireless localization and MIMO channel prediction.

\textbf{Channel Estimation}. Channel estimation is a critical task in wireless systems, with past works employing training pilots~\cite{marzetta2006fast}, feedback~\cite{fan2020towards}, parametric or empirical models~\cite{vasisht2016eliminating}, and blind or opportunistic methods~\cite{ma2018enabling}. As the number of antennas skyrockets, the classic feedback-based estimation methods bring excessive overhead~\cite{vasisht2016eliminating}. Recent works~\cite{liu2021fire,bakshi2019fast,huang2019deep,liaskos2020end} attempt to use deep learning to fill out the gap. NNCONFIG~\cite{liaskos2020end} proposed a scheme for mapping the trained neural networks to intelligent surfaces. State-of-the-art work, FIRE~\cite{liu2021fire}, proposes a generative model based on VAE to learn the downlink channel from the uplink feedback. However, they are purely data-driven machine learning models and highly rely on massive data. Unlike them, \oursystem incorporates a physical model (RF radiance field) into the learning process, enhancing interpretability and channel learning accuracy through prior wave transmission knowledge.

\textbf{Wireless Localization}. Wireless localization is a long-studied topic with extensive works~\cite{ni2003landmarc,yang2013rssi,yang2014tagoram}. Past works locate a device by building the transmission model between the position and various metrics of received RF signals: RSSI~\cite{ni2003landmarc}, phase~\cite{ma2017drone,yang2014tagoram}, CSI~\cite{yang2013rssi,xie2019md}, Time-of-Flight~\cite{mariakakis2014sail}, and AoA~\cite{an2020general,ayyalasomayajula2018bloc,xiong2013arraytrack,xie2018swan}. They are widely used in Wi-Fi~\cite{kotaru2015spotfi,ayyalasomayajula2020deep,xie2018swan}, Bluetooth~\cite{ayyalasomayajula2018bloc,cominelli2019dead}, RFID~\cite{an2020general,wang2014rf,wang2017d}, and LoRa~\cite{bnilam2020loray}. However, classic geometric or empirical localization models suffer from problems caused by the environment (e.g., complex multipath effect, non-line of sight, etc. These problems are extensively studied in literature~\cite{wang2013dude,ayyalasomayajula2020deep}. Recent state-of-the-art approaches leverage deep learning models to enhance accuracy in complex indoor scenarios~\cite{ayyalasomayajula2020deep,an2020general,zheng2019zero,raza2019dataset} but require extensive training data, limiting real-world applications. In contrast, \oursystem embeds physical wave transmission knowledge into learning networks, improving model efficiency through turbo-learning and reducing raw training data while maintaining localization accuracy, thus offering a more practical solution.

\section{Conclusion and Future Work}

We present \oursystem, a novel deep learning architecture for wireless channel understanding. \oursystem first embeds the entire electromagnetic wave transmission physical model into the channel learning model. Our extensive experiments show that \oursystem can boost the performance of many deep-learning based application-layer tasks, including indoor localization and massive MIMO communication. Future work will focus on addressing the following challenges and improvements:

\textbf{Generalizability}. Transferring \oursystem across different scenes remains a substantial challenge. Potential solutions include pre-training on large datasets from multiple scenes for improved generality, employing incremental learning to adapt to surrounding noise or dynamic environments, and utilizing transfer learning to maximize model reusability across various scenes.

\textbf{Time consumption.} \oursystem needs several hours to train the model. However, recent substantial research has been devoted to reducing the time by compression methods. For example, Instant-NGP~\cite{muller2022instant} can train an optical NeRF in 5 seconds. Consequently, future work will explore optimizing training time consumption through these methods.

\begin{acks}
This study is supported by NSFC Key Program (No. 61932017), NSFC Excellent Young Scientists Fund (Hong Kong and Macau) (No. 62022003), NSFC General Program (No. 61972331), and UGC/GRF (No. 15204820, 15215421). We thank all the anonymous reviewers and the shepherd, for their valuable comments and helpful suggestions.
\end{acks}

\newpage
{\small
\bibliographystyle{IEEEtran}
\bibliography{nerf2.bib}
}

\end{document}